\documentclass[aps,prb,reprint,groupedaddress,showpacs,dvipdfmx]{revtex4-1} 
\usepackage{graphicx}
\usepackage{dcolumn}
\usepackage{bm}
\usepackage{amsmath}
\usepackage{amssymb}
\usepackage{txfonts}
\usepackage{url}
\usepackage{multirow}
\usepackage{here}
\usepackage[usenames,dvipsnames]{color}
\definecolor{Gray}{gray}{0.0}
\definecolor{lightGray}{gray}{0.35}

\makeatletter
\def\Hline{%
\noalign{\ifnum0=`}\fi\hrule \@height 1pt \futurelet
\reserved@a\@xhline}
\makeatother

\begin{document}
\title{All--electron quantum Monte Carlo with Jastrow single determinant Ansatz: application to the sodium dimer}
\author{Kousuke Nakano$^{1,2}$}
\email{kousuke\_1123@icloud.com}

\author{Ryo Maezono$^{2,3}$}

\author{Sandro Sorella$^{1}$}
\email{sorella@sissa.it}
 
\affiliation{$^{1}$
 International School for Advanced Studies (SISSA), Via Bonomea 265, 34136, Trieste, Italy}
 
\affiliation{$^{2}$
 School of Information Science, Japan Advanced Institute of Science and Technology (JAIST), Asahidai 1-1, Nomi, Ishikawa
 923-1292, Japan}
 
\affiliation{$^{3}$
 Computational Engineering Applications Unit, 
 RIKEN, 2-1 Hirosawa,
 Wako, Saitama 351-0198, Japan}

\date{\today}
\begin{abstract}
In this work, we report potential energy surfaces (PESs) of the sodium dimer calculated by variational (VMC) and lattice regularized diffusion Monte Carlo (LRDMC). 
The VMC calculation is accurate for determining the equilibrium distance and the qualitative shape of the experimental PES. 
Remarkably, after the application of the LRDMC projection to this single determinant ansatz, namely the Jastrow Antisymmetrized Geminal Product (JAGP), chemical accuracy ($\sim$ 1kcal/mol) is reached, and the obtained dissociation energy, equilibrium internuclear distance, and harmonic vibrational frequency are in very good agreement with the experimental ones. 
This outcome crucially depends on the quality of the optimization used to determine the best possible trial function within the chosen ansatz. 
The strategy adopted in this work is to minimize the variational energy by initializing the trial function with the DFT single determinant ansatz expanded exactly in the same atomic basis used for the corresponding VMC and LRDMC calculations. 
This atomic basis is ad-hoc reshaped for QMC calculations. 
Indeed, we multiply the standard Gaussian type atomic orbitals by a one-body Jastrow factor, satisfying in this way the electron-ion cusp conditions. 
This allows us to use a very small basis almost converged in the complete basis set limit, by reducing the computational effort as well as the statistical fluctuations on the total energy.
In order to achieve these important advantages, we have defined a very efficient DFT algorithm in the mentioned basis, by estimating the corresponding matrix elements on a mesh, and by using a much finer mesh grid in the vicinity of nuclei.

\end{abstract}
\maketitle


\section{Introduction}
First principle quantum Monte Carlo (QMC) techniques, such as variational quantum Monte Carlo (VMC) and diffusion quantum Monte Carlo (DMC) are among the state-of-art numerical methods used to obtain highly accurate many-body wavefunctions{~\cite{2001FOU}}. 
Recent developments in QMC enable us to calculate not only the ground state energy but also vibrational frequencies{~\cite{2014LUO, 2013ZEN}} and excited states{~\cite{2018MUS, 2018HUN}}, as well as to study phase diagrams of materials{~\cite{2018MAZ}} and determine quantitative properties of a metal-insulator transition{~\cite{2011STE}} or an excitonic behavior{~\cite{2017VAR}}. 
Although QMC has been mainly applied to model compounds such as atoms or small molecules due to the large computational cost, it should be much more feasible for "real materials" (e.g. protein, surface, glass, etc.) in the near future, because the QMC algorithm scales 
very well with the number $N$ of electrons -at most $N^4$- and sustains almost ideal scaling in massively parallel architectures{~\cite{2001FOU}}.

\vspace{2mm}
In order to apply QMC for "real materials", it is convenient to replace core electrons with pseudopotentials because they have a little effect on chemical properties, and their replacement can reduce the QMC computational cost~{\cite{1986CEP, 1987HAM, 2005MA}} by a factor proportional to Z$^{5.5 \sim 6.5}$, where Z is the atomic number.
Nevertheless, all-electron calculations are important as they represent 
useful benchmarks for highly accurate methods, removing the problem to 
work with a very accurate pseudopotential, though several 
progress have been made recently{~\cite{2015TRA, 2016KRO, 2017TRA, 2017BEN, 2018BEN, 2018ANN}}.
Unfortunately, within QMC, all-electron calculations are rarely applied for atoms of large atomic number, mainly because they are too much computationally demanding at least in the simplest formulation of the VMC and DMC algorithms. 
Indeed, some progress has been obtained in VMC by considering more sophisticated 
trial moves in the Metropolis algorithm. Umrigar et al.{\cite{1993UMR, 1998STE}} have proposed an accelerated Metropolis method to reduce fluctuations in VMC of full-core atoms, wherein electrons close to the atomic cores are displaced much more slowly than those in the valence region. 
Analogously in DMC {~\cite{2001FOU}} the time step is decreased only around nuclei, by improving the efficiency of the algorithm as compared with a standard all-electron calculation with a single very small time step. Moreover, a very accurate trial wave function is necessary for a reasonably efficient QMC all-electron simulation because otherwise large absolute values of kinetic and potential energies around nuclei usually induce large statistical fluctuations. For the latter problem, in this work, we determine the trial function by using an appropriate density-functional theory (DFT) algorithm working in the atomic basis set that exactly satisfies the electron-ion cusp condition. Indeed, we have experienced that the large fluctuations in the corresponding VMC and DMC calculations can be substantially reduced by adopting a single determinant trial wave function that is particularly 
accurate in the vicinity of the nuclei.

\vspace{3mm}
In order to obtain a good trial wave function, 
the Kohn-Sham Hamiltonian matrix elements involving a rapidly varying electron density (wavefunction) in the vicinity of nuclei should be accurately evaluated in DFT.
The matrix elements are composed of integrals of the kinetic, the Coulomb (Hartree and electron-ion), and the exchange-correlation (XC) parts{~\cite{2004MAR}}.
When the wave function is expanded using Gaussian-type orbitals (GTOs), integrals of the kinetic and the Coulomb parts can be done analytically{~\cite{1996SZA}}.
However, the integral should be done numerically if the wave function is expanded in Slater-type orbitals (STOs), 
as well as in our modified GTO basis.
In this case, Poisson's equation is solved to determine the Hartree potential by integrating the Coulomb kernel over the electron density calculated at each point of the mesh. This scheme is employed in our TurboRVB code{~\cite{2019SOR}}.
If Poisson's equation is solved in real space (e.g. the finite element method{~\cite{2000BEC}}),
the integral can be evaluated without no other approximation than the finite mesh. Moreover, an arbitrary fine mesh grid can be used in the vicinity of nuclei, within the so-called multigrid approach{~\cite{2000BEC}}.
On the other hand, the multigrid approach is not easily implemented when the very efficient 
fast Fourier transform (FFT) 
is used to solve Poisson's equation. In this case, the grid should be {\it uniform} in all the space, which increases a computational cost. 
This drawback can be solved by the so-called pseudo-charge method within the LAPW technique{~\cite{1994SIN, 2004MAR}}, but the corresponding implementation is rather involved. 
In this work, we determine the Hartree potential with standard FFT convolution on a coarse mesh, 
then interpolate these values on a much finer mesh in the vicinity of nuclei. 
We show that this is enough to determine a good trial function that can be used as a suitable starting guess for QMC energy optimization.
Although the DFT energy obtained with the above approximation is not exactly consistent with the one corresponding to a very dense uniform mesh, the VMC energies and the variances of the obtained initial trial wave functions are almost indistinguishable each other. We emphasize that this is just due to the simple and efficient interpolation scheme of the Hartree potential that we have introduced in this work. 

\vspace{3mm}
This method is applied to the sodium dimer, which has been extensively studied both experimentally{~\cite{1909WOO, 1929FRE, 1969DEM, 1975DEM, 1978KUS, 1983VER, 1997KAM, 2007BAI}} and theoretically{~\cite{1993MAG,2000HO,2004MAT,2004MARO,2005HAR,2007BEN,2011MUS,2012MUS,2012MOR}} in the past decades.
Several all-electron VMC and DMC studies have been reported for various atoms and molecules,
{\cite{1982REY, 1986REY, 1988VRB, 1990LES, 1993UMR, 1995KEN, 1996LUC, 1997YOS, 1997HUA, 1999SHL,
2002SAR, 2003CAS, 2004CAS, 2005MA, 2005CAF, 2006BUE, 2010NEM, 2014SCE, 2016POW}}
but, to our knowledge, only one paper has reported the sodium dimer{~\cite{2010NEM}}, wherein the dissociation energy at the experimental equilibrium distance has been calculated.
Moreover, the full potential energy surface (PES) and other spectroscopic properties such as harmonic vibration frequency have not been calculated using all-electron VMC and DMC yet. All-electron calculations for the sodium dimer is informative as a reference because it is known that the use of a pseudopotential sometimes induces discrepancy due to the presence of the semi-core electrons{~\cite{2003MAE}}. 
We successfully calculated PESs of the sodium dimer with small statistical errors, and the obtained dissociation energy, equilibrium internuclear distance, and harmonic vibrational frequency are in very good agreement with the experimental values.
The main outcome of this work is that, after the optimization of the energy, a single determinant ansatz, the so-called JAGP described in the next section, 
can accurately describe this very weak and challenging chemical bond, within QMC technique.
This is very important because a single determinant ansatz can be extended to much larger systems, even within the computationally demanding QMC methods. On the contrary, the multireference approach would be certainly impossible in this case, because it requires a number of determinants exponentially large in the number of electrons, and 
a corresponding computational burden.


\section{Methodology}

\subsection{Variational and Lattice regularized Diffusion Monte Carlo}
The Jastrow single determinant Ansatz, a Jastrow-Slater determinant (JSD) and Jastrow antisymmetrized geminal power (JAGP){~\cite{2003CAS}} variational wave functions, are defined by the product of two terms, namely a Jastrow
$J$ and an antisymmetric part ($\Psi = J{\Psi _\text{AGP/SD}}$).
The Jastrow term is composed of one-body, two-body and three/four-body factors ($J = {J_1}{J_2}{J_{3/4}}$).
The one-body and two-body factors are used to fulfill the electron-ion and electron-electron cusp conditions, respectively. The one-body Jastrow factor reads:
\begin{equation}
{J_1}\left( {{{\vec r}_1}, \ldots {{\vec r}_N}} \right) = \exp \left( {\sum\limits_{i,I,l} {g_{I,l}^{}\chi _{I,l}^J\left( {{{\vec r}_i}} \right)} } \right) \cdot \prod\limits_i {{{\tilde J}_1}\left( {{{\vec r}_i}} \right)},
\end{equation}
\begin{equation}
{\tilde J_1}\left( {\vec r} \right) = \exp \left( {\sum\limits_I { - {{\left( {2{Z_I}} \right)}^{3/4}}u\left( {{2Z_I}^{1/4}\left| {\vec r - {{\vec R}_I}} \right|} \right)} } \right),
\label{onebody_j_single}
\end{equation}
where ${\vec r_i}$ are the electron positions, $R_I$ are the atomic positions with corresponding atomic number $Z_I$, $l$ runs over single-particle orbitals ${\chi _{I,l}^J}$ centered on the atom $I$, and ${u\left( r \right)}$ contains a variational parameter $b$:
\begin{equation}
u\left( r \right) = \frac{b}{2}\left( {1 - {e^{ - r/b}}} \right).
\label{onebody_u}
\end{equation}
The two-body Jastrow factor reads:
\begin{equation}
{J_2}\left( {{{\vec r}_1}, \ldots {{\vec r}_N}} \right) = \exp \left( {\sum\limits_{i < j} {v\left( {{r_{i,j}}} \right)} } \right),
\end{equation}
where ${r_{i,j}} = \left| {{{\vec r}_i} - {{\vec r}_j}} \right|$ is the distance between two electrons, and $v\left( r \right)$ contains a variational parameter $F$:
\begin{equation}
v\left( r \right) = \frac{r}{2}{\left( {1 - F\,r} \right)^{ - 1}}.
\end{equation}
The three-body Jastrow factor reads:
\begin{equation}
{J_{3/4}}\left( {{{\vec r}_1}, \ldots {{\vec r}_N}} \right) = \exp \left( {\sum\limits_{i < j} {{\Phi _J}\left( {{{\vec r}_i},{{\vec r}_j}} \right)} } \right)
\end{equation}
\begin{equation}
{\Phi _J}\left( {{{\vec r}_i},{{\vec r}_j}} \right) = \sum\limits_{l,m,a,b} {g_{l,m}^{a,b}\chi _{a,l}^J\left( {{{\vec r}_i}} \right)\chi _{b,m}^J\left( {{{\vec r}_j}} \right)},
\end{equation}
where the indices $l$ and $m$ again indicate different orbitals centered on
corresponding atoms $a$ and $b$. 
In the present study, the coefficients of the three/four-body
Jastrow factor were set to zero in case of $a \ne b$.
The antisymmetric part reads:
\begin{equation}
{\Psi _\text{AGP}}\left( {{{\vec r}_1}, \ldots ,{{\vec r}_N}} \right) = \det \left( {{\Phi _\text{AGP}}\left( {{{\vec r}_i},{{\vec r}_j}} \right)} \right)
\end{equation}
and the geminal function is expanded over an atomic basis:
\begin{equation}
{\Phi _\text{AGP}}\left( {{{\vec r}_i},{{\vec r}_j}} \right) = \sum\limits_{l,m,a,b} {\lambda _{a,b}^{l,m}{\phi _{a,l}}\left( {{{\vec r}_i}} \right){\phi _{b,m}}\left( {{{\vec r}_j}} \right)}
\end{equation}
where indices $l$ and $m$ indicate different orbitals centered on
atoms $a$ and $b$, and $i$ and $j$ are coordinates of spin up and down
electrons, respectively.
The antisymmetric part can also be represented by molecular orbitals{~\cite{2004CAS}}:
\begin{equation}
{\Phi _\text{AGP}}\left( {{{\vec r}_i},{{\vec r}_j}} \right) = \sum\limits_k^L {{\lambda _k}{{\tilde \psi }_k}\left( {{{\vec r}_i}} \right){{\tilde \psi }_k}\left( {{{\vec r}_j}} \right)},
\end{equation}
\begin{equation}
{\tilde \psi _k}\left( {\vec r} \right) = \sum\limits_a^M {\sum\limits_l^{{L_a}} {c_{a,l}^k{\phi _{a,l}}\left( {\vec r} \right)} },
\end{equation}
where $M$ is the number of atoms, ${{L_a}}$ is the number of atomic orbitals belonging to atom $a$, ${c_{a,l}^k}$ are the coefficients of the atomic orbitals, and $L$ is the number of molecular orbitals. If $L$ is equal to the half of the total number of electrons ($N/2$), the antisymmetric part coincides with the Slater determinant{~\cite{2003CAS, 2004CAS}}.
In this study, the cc-pDVZ basis set taken from EMSL Basis Set Library{~\cite{1996FEL,2007SCU}} was used for the atomic orbitals ${\phi _{a,l}}\left( {{{\vec r}_i}} \right)$ of sodium both in JSD and JAGP ansatz. According to the scheme recently introduced by Mazzola et. al.~{\cite{2018MAZ}}, the orbitals whose exponents ($Z$) are larger than 300 were cut to avoid numerical instabilities.
The modified cc-pDVZ basis for the sodium was finally composed of 8s7p1d for the determinant part. 
The basis set was treated as an uncontracted one and the exponents are relaxed in the optimization procedure.
We note that the large exponent elements removed from the basis set are taken into account implicitly by means of the one-body Jastrow term~{\cite{2018MAZ}} that indeed allows us to fulfill exactly the electron-ion cusp conditions.
The variational JSD and JAGP wave functions were optimized using the stochastic configuration in combination with the linear method~{\cite{benzene,linear}} that enable us to optimize thousands of parameters simultaneously even within 
a stochastic optimization technique.


\vspace{3mm}
Lattice regularized diffusion Monte Carlo (LRDMC) 
is a projection technique that allows a systematic improvement of the variational ansatz, yielding the corresponding one with the lowest energy 
and the same signs in configuration space. This energy is the so-called 
"fixed-node" DMC energy and can be obtained 
 with the standard short time discretization\cite{1993UMR}- i.e. the conventional approach- or by the so-called lattice regularization, namely by discretizing on 
 a lattice the continuous Hamiltonian{~\cite{1994TEN, 1998BUO, 2000SOR}}.
We summarize the method here by emphasizing some important improvements for the all-electron case studied here. The interested readers should refer to Refs. {\onlinecite{2005CAS, 2006CAS, 2010CAS, 2017BEC}} for details.
In LRDMC, the original continuous Hamiltonian is regularized by an approximate one ${H^a}$, such that ${H^a} \to H$ for $a \to 0$,
where $a$ is the parameter used to discretize the continuous space.
We consider the Hamiltonian in atomic units:
\begin{equation}
H = - \frac{1}{2}\sum\limits_i^N {{\Delta _i} + V\left( {\vec x} \right)} + \sum\nolimits_{I < J} {\frac{{{Z_I}{Z_J}}}{{\left| {{{\vec R}_I} - {{\vec R}_J}} \right|}}},
\end{equation}
where $N$ is the number of electrons, $\vec x$ is 3$N$ dimension electron coordination,
$\vec x = \left( {{{\vec r}_1},{{\vec r}_2}, \ldots ,{{\vec r}_N}} \right)$, and 
$V\left( {\vec x} \right) = {V_\text{ee}}\left( {\vec x} \right) + {V_\text{ei}}\left( {\vec x} \right)$ is the standard many-body potential, which includes the electron-electron interaction:
\begin{equation}
{V_\text{ee}}\left( {\vec x} \right) = \sum\nolimits_{i < j} {\frac{1}{{\left| {{{\vec r}_i} - {{\vec r}_j}} \right|}}},
\end{equation}
and the electron-ion interaction:
\begin{equation}
{V_\text{ei}}\left( {\vec x} \right) = - \sum\limits_i^{} {{\nu _\text{ei}}\left( {{{\vec r}_i}} \right)} = - \sum\limits_i^{} {\sum\limits_I^{} {\frac{{{Z_I}}}{{\left| {{{\vec r}_i} - {{\vec R}_I}} \right|}}} },
\end{equation}
where ${{{\vec R}}}$ and ${{{\vec r}}}$ are the ionic and electron positions, respectively.
The kinetic part is approximated by a finite difference form:
\begin{equation}
{\Delta _i} \approx \Delta _i^a = \Delta _i^{a,p} + \Delta _i^{a',1 - p},
\end{equation}
where $\Delta _i^{a,p}$ is defined by a mesh size $a$ and a function $p\left( {\vec r} \right)$:
\begin{equation}
\begin{split}
\Delta _i^{a,p}f\left( {{x_i},{y_i},{z_i}} \right) &= \frac{1}{{{a^2}}}[p\left( {{x_i} + a/2} \right)\left( {f\left( {{x_i} + a} \right) - f\left( {{x_i}} \right)} \right) \\
&+ p\left( {{x_i} - a/2} \right)\left( {f\left( {{x_i} - a} \right) - f\left( {{x_i}} \right)} \right)] + {x_i} \leftrightarrow {y_i} \leftrightarrow {z_i}.
\end{split}
\end{equation}
In this work, we have adopted a more convenient and simpler form for the 
the function $p\left( {\vec r} \right)$ that is chosen as:
\begin{equation}
p\left( {\vec r} \right) = \exp \left( { - 4{{\left| {\vec r - {{\vec R}_\text{c}}} \right|}^2}} \right),
\end{equation}
where ${{\vec R}_\text{c}}$ is the position of the nucleus closest to the electron in $\vec r$. The function $p$ decays much faster than the original form ( $p(\vec r) = 1/\left( {1 + {Z^2}/4{{\left| {\vec r - {{\vec R}_{\rm{c}}}} \right|}^2}} \right)$ ) and enables us to use the larger lattice space $a'$ 
in a wider valence region, because the small lattice space $a$ is used only if the electron is very close to the nucleus.
The constant $a'/a$ is set to an irrational number in order to sample all the 
continuous space of the original Hamiltonian\cite{2005CAS}.
The potential term $V\left( x \right) = {V_\text{ee}}\left( x \right) + {V_\text{ei}}\left( x \right)$ is also discretized by the parameter $a$ to realize a smooth convergence for $a \to 0$. The electron-electron potential is not necessarily discretized, but the electron-ion one is modified as:
\begin{equation}
{\nu _\text{ei}}\left( {{{\vec r}_i}} \right) \to \nu _{\max ,i}^a\left( {\vec x} \right) = {\rm{Max}}\left[ {\nu _{\text{zv},i}^a\left( {\vec x} \right),{\nu _\text{ei}}\left( {{{\vec r}_i}} \right)} \right],
\label{max_ele_ion}
\end{equation}
\begin{equation}
\nu _{\text{zv},i}^a\left( {\vec x} \right) = {\nu _\text{ei}}\left( {{{\vec r}_i}} \right) + \frac{{\left( {\nabla _{i,a}^2 - \nabla _i^2} \right){\Psi _\text{G}}\left( {\vec x} \right)}}{{2{\Psi _\text{G}}\left( {\vec x} \right)}},
\end{equation}
where ${\Psi _\text{G}}\left( {\vec x} \right)$ is a guiding function, and $\nabla _{i,a}^2 = \Delta _i^a$.
Although the electron-ion potential (${\nu _\text{ei}}\left( {{{\vec r}_i}} \right)$) in the right-hand side of Eq.{~(\ref{max_ele_ion})} was regularized by:
\begin{equation} \label{unreg}
{\nu _\text{ei}}\left( {{{\vec r}_i}} \right) = \sum\limits_I^{} {\frac{{{Z_I}}}{{\text{Max}\left( {\left| {{{\vec r}_i} - {{\vec R}_I}} \right|,a} \right)}}}
\end{equation}
to cut the Coulomb singularity at small distances, we have noticed that this regularization is not necessary within the so-called fixed-node approximation during this work. This is because $\nu _{\text{zv},i}^a\left( {\vec x} \right)$ does not diverge even when the electron-ion distance is small. Therefore, the Eq.{~(\ref{max_ele_ion})} ensures the removal of the singularity unless in the vicinity of the nodal surface. The fixed node approximation removes also this singularity and
therefore the algorithm remains always stable even without the use of Eq.(\ref{unreg}).
Now that the Hamiltonian is discretized, the efficient lattice Green function Monte Carlo algorithm{~\cite{1994TEN, 1998BUO, 2000SOR}}, which is valid on a lattice model, can be applied straightforwardly:
\begin{equation}
\left\langle x \right|{H^a}\left| {{\Psi _\text{G}}} \right\rangle = \sum\limits_{x'}^{} {H_{x,x'}^a\left\langle {{x'}}
 \mathrel{\left | {\vphantom {{x'} {{\Psi _\text{G}}}}}
 \right. \kern-\nulldelimiterspace}
 {{{\Psi _\text{G}}}} \right\rangle }
\end{equation}
The resulting algorithm is called LRDMC.
The corresponding Green function matrix elements with the important sampling 
are ${G_{x,x'}} = {\Psi _\text{G}}\left( {\vec x'} \right)\left( {{\Lambda _{x,x'}} - H_{x',x}^a} \right)/{\Psi _\text{G}}\left( {\vec x} \right)$, where $\Lambda$ is a diagonal matrix with ${{\Lambda _{x,x'}}} = \lambda$ and $\lambda$ should be a sufficiently large to obtain the ground state.
The LRDMC algorithm is as follows{~\cite{2005CAS}}: given a walker with configuration ${\vec x}$ and weight $w$, 
a new configuration $x' \ne x$ is ${p_{x,x'}} = {G_{x,x'}}{\rm{ /}}{b_x}$, where ${b_x} = \sum\nolimits_{x' \ne x} {{G_{x',x}}}$ is a normalization factor to make the Green function a transition probability.
The walker weight is updated by a factor $w \to w\exp \left( { - {\tau _x}{e_L}\left( {\vec x,\left[ {{\Psi _\text{G}}} \right]} \right)} \right)$, where ${\tau _x} = - \log \left( {rand} \right)/{b_x}$ is a diffusion time determined by a random number $0 < rand \le 1$, and ${e_L}\left( {\vec x,\left[ {{\Psi _\text{G}}} \right]} \right) = \sum\limits_{x'}^{} {H_{x',x}^a{\Psi _\text{G}}\left( {\vec x'} \right)} /{\Psi _\text{G}}\left( {\vec x} \right)$ is a local energy with the guiding function. Of course, the usual branching scheme and many walker technique can also be used. 
Unfortunately, the Green function cannot be made strictly positive for fermions, therefore, the fixed-node approximation 
should be introduced{~\cite{2017BEC}}. To avoid the sign problem, the Hamiltonian is modified using the spin-flip term ${\mathcal{V}_{{\rm{sf}}}}\left( {\vec x} \right) = \sum\limits_{x':{s_{x,x'}} > 0}^{} {{H_{x,x'}}{\Psi _\text{G}}\left( {\vec x'} \right)} /{\Psi _\text{G}}\left( {\vec x} \right)$:
\begin{equation}
H_{x,x'}^{\text{FN},\gamma } = \left\{ \begin{array}{l}
{H_{x,x}} + \left( {1 + \gamma } \right){\mathcal{V}_{{\rm{SF}}}}\left( {\vec x} \right)\,\,\,\,{\rm{for}}\,\,\,\,x' = x,\\
{H_{x,x'}}\,\,\,\,\,\,\,\,\,\,\,\,\,\,\,\,\,\,\,\,\,\,\,\,\,\,\,\,\,\,\,\,\,\,\,\,\,\,\,\,{\rm{for}}\,\,\,\,x' \ne x,{s_{x,x'}} < 0,\\
 - \gamma {H_{x,x'}}\,\,\,\,\,\,\,\,\,\,\,\,\,\,\,\,\,\,\,\,\,\,\,\,\,\,\,\,\,\,\,\,\,{\rm{for}}\,\,\,\,x' \ne x,{s_{x,x'}} > 0,
\end{array} \right.
\end{equation}
where ${s_{x,x'}} = {\Psi _\text{G}}\left( {\vec x} \right){H_{x,x'}}{\Psi _\text{G}}\left( {\vec x'} \right)$ and $\gamma \ge 0$ is a real parameter. Finally, a mixed average of the fixed-node Hamiltonian:
\begin{equation}
{E_{{\rm{MA}}}} = \frac{{\left\langle {{\Psi _\text{G}}} \right|H_\text{FN}^a\left| {{\Upsilon _0}} \right\rangle }}{{\left\langle {{{\Psi _\text{G}}}} \mathrel{\left | {\vphantom {{{\Psi _\text{G}}} {{\Upsilon _0}}}} \right. \kern-\nulldelimiterspace} {{{\Upsilon _0}}} \right\rangle }}
\end{equation}
can be calculated by the weights and local energies after sufficient number of projections, where ${\left| {{\Upsilon _0}} \right\rangle }$ is the ground-state wave function. It has been confirmed that the mixed average energy is consistent with the fixed-node energy of the standard DMC in the limit $a \to 0${~\cite{2005CAS}}.


\subsection{DFT algorithm in the same basis used for QMC}
The trial functions for the Jastrow single determinant ansatz were determined from 
DFT calculations by using a single determinant ansatz expanded exactly in the same atomic 
basis used for the corresponding VMC and LRDMC calculations. 
Within DFT, the Hamiltonian and the overlap matrix elements required for the solution of the Kohn-Sham equations are represented as:
\begin{equation}
H_{i,j}^{a,b} = \int_{}^{} {d\vec r\tilde \phi _j^b\left( {\vec r - {{\vec R}_b}} \right)\hat H\left( {\vec r} \right)\tilde \phi _i^a\left( {\vec r - {{\vec R}_a}} \right)},
\end{equation}
\begin{equation}
S_{i,j}^{a,b} = \int_{}^{} {d\vec r\tilde \phi _j^b\left( {\vec r - {{\vec R}_b}} \right)\tilde \phi _i^a\left( {\vec r - {{\vec R}_a}} \right)},
\end{equation}
where $\tilde \phi _i^a\left( {\vec r - {{\vec R}_a}} \right)$ and $\tilde \phi _j^b\left( {\vec r - {{\vec R}_b}} \right)$ are $i$-th, $j$-th GTO for atom $a$ and $b$ multiplied by one-body Jastrow factor, namely:
\begin{equation}
\tilde \phi _j^b\left( {\vec r - {{\vec R}_b}} \right) = \phi _j^b\left( {\vec r - {{\vec R}_b}} \right){\tilde J_1}\left( {\vec r} \right),
\end{equation}
where ${{\tilde J}_1}\left( {\vec r} \right)$ is the same as in Eq.~(\ref{onebody_j_single}).
This formulation allows us to use a very small basis almost converged in the complete basis set limit, by reducing in this way the computational effort as well as the statistical fluctuations 
on the total energy{~\cite{2018MAZ}}.
Indeed, as it is simple to show, each element of the basis set satisfies the 
so-called electron-ion cusp condition, namely that when $ \vec r $ is close 
to any atomic position ${\vec R}_b$:
\begin{equation}
 \lim\limits_{\vec r \to {\vec R}_b} { \nabla \tilde \phi_j^a \over \tilde \phi_j^a } = 
 -Z_b {\vec r-{\vec R}_b \over |\vec r -{\vec R}_b| } 
\end{equation}
 for all $a,b$.
\vspace{3mm}
In order to construct the trial wave function efficiently, we have defined an efficient DFT 
algorithm in the mentioned basis, by estimating the corresponding matrix elements on a mesh, 
and by using a much finer mesh grid in the vicinity of nuclei.
The Hamiltonian operator is composed of:
\begin{equation}
\hat H\left( {\vec r} \right) = \hat T + {V_\text{H}}\left( {\vec r} \right) + {V_\text{ext}}\left( {\vec r} \right) + {V_\text{XC}}\left( {\vec r} \right),
\end{equation}
where $\hat T$ is a kinetic operator, ${V_\text{H}}\left( {\vec r} \right)$ is the Hartree (electron-electron) potential, ${V_\text{ext}}\left( {\vec r} \right)$ is the electron-ion potential (that may or may not include a true external potential)~ and ${V_\text{XC}}\left( {\vec r} \right)$ is the exchange-correlation potential.
Given that the wavefunction is expanded in atomic orbitals such as GTO, the kinetic, electron-ion, and exchange-correlation terms are readily calculated at {\it any point in real space}.
On the other hand, the Hartree potential is not determined in this manner, since it can be evaluated more conveniently on a uniform grid by solving Poisson's equation with FFT:
\begin{equation}
\nabla {V_\text{H}}\left( {\vec r} \right) = - 4\pi \rho \left( {\vec r} \right),
\end{equation}
where $\rho \left( {\vec r} \right)$ is the charge density.
Therefore, the use of FFT with a fine grid in the vicinity of nuclei {\it necessarily} involves the same fine grid in the interstitial regions where the electron density smoothly changes, which gratuitously increases the computational cost.
In our implementation, we have found a good compromise between accuracy and efficiency in the following way. 
The Hartree potential is calculated first on a {\it coarse uniform } grid by solving Poisson's equation with the FFT algorithm. In the second step, the Hartree potential is interpolated on a {\it fine grid} only in the vicinity of nuclei using standard interpolation methods such as linear or cubic.
A schematic figure of the linear interpolation in the two-dimensional case is shown in Fig.{~\ref{Figure1}}. The values on the fine grid are interpolated using the nearest four points, namely: 
\begin{eqnarray}
{V_\text{H}}\left( {{x_i},\,\,{y_j}} \right) &=& \left( {1 - \frac{{{s_x}}}{n}} \right) \cdot \left( {1 - \frac{{{s_y}}}{n}} \right) \cdot {V_\text{H}}\left( {{X_I},\,\,{Y_J}} \right) \\ \nonumber
 &+& \frac{{{s_x}}}{n} \cdot \left( {1 - \frac{{{s_y}}}{n}} \right) \cdot {V_\text{H}}\left( {{X_{I + 1}},\,\,{Y_J},} \right)\\ \nonumber
 &+& \left( {1 - \frac{{{s_x}}}{n}} \right) \cdot \frac{{{s_y}}}{n} \cdot {V_\text{H}}\left( {{X_I},\,\,{Y_{J + 1}}} \right)\\ \nonumber
 &+& \frac{{{s_x}}}{n} \cdot \frac{{{s_y}}}{n} \cdot {V_\text{H}}\left( {{X_{I + 1}},\,\,{Y_{J + 1}}} \right)
\end{eqnarray}
where ${X_{I}}$ and ${Y_{J}}$ represent coarse grid points, 
$x_{i = nI + {s_x}}$ and $y_{j = nJ + {s_y}}$ represent interpolated points in the vicinity of nuclei, $n$ is the ratio of the interpolated grid to the coarse one (i.e $n-1$ coincides with the number of interpolated fine points between the coarse-grid ones), and $0 \le {s_x}, {s_y} < n$. The values on the fine grid is interpolated by the nearest eight points in three dimensional case. The cubic interpolation is performed using the nearest twenty-four points.
As a result, the matrix elements of Hamiltonian can be evaluated by combining a coarse grid and an interpolated fine grid in the vicinity of nuclei.
Notice that a similar interpolation was done in pseudopotential calculation{~\cite{1999ONO, 2005ONO}}, wherein the interpolation scheme was used to evaluate inner products between wave functions and nonlocal parts of pseudopotentials.
The total DFT energy corresponding to the chosen interpolation 
for the Hartree potential is sizably different from the one 
obtained with a very fine grid (namely converged).
However, VMC and LRDMC energies obtained with the Kohn-Sham Slater determinants with or without the interpolation scheme proposed, are very close, 
indicating that our DFT scheme provides very good Kohn-Sham orbitals, despite the observed error in the DFT energy.


\section{Validation of the interpolation scheme}
To investigate the quality of the trial wave functions obtained by the interpolation technique, 
ground state energies of the sodium atom were calculated using DFT, VMC, and LRDMC.
DFT calculations were performed with a single fine grid or using the interpolation scheme, wherein the LDA functional developed by Perdew and Zunger{~\cite{1981PER}} was employed.
Three types of single-grid DFT calculations were performed with
(0.02 Bohr)$^3$, (0.05 Bohr)$^3$, and (0.10 Bohr)$^3$ grids.
For comparison, three types of DFT calculations using the cubic interpolation method were performed, namely, (0.01 Bohr)$^3$ grid was used for the core electron region, while (0.05 Bohr)$^3$, (0.10 Bohr)$^3$ or (0.20 Bohr)$^3$ grids were used for the valence electron region, wherein the core-electron region, centered on the sodium atom, has a volume of (2.00 Bohr)$^3$. The calculation using the linear interpolation method was also performed using (0.01 Bohr)$^3$ and (0.20 Bohr)$^3$ grids.
Then, three types of VMC and LRDMC calculations were performed starting from the resultant wave functions. VMC-JDFT denotes that only the Jastrow factor was optimized using the Jastrow-Slater ansatz, namely, the nodal surface was determined by the DFT.
VMC-JSD and VMC-JAGP denote that both Jastrow and determinant parts were optimized using Jastrow-Slater and Jastrow antisymmetrized geminal power ansatz, respectively.
LRDMC (GF=JDFT, JSD, JAGP) denotes that the wave functions optimized using each ansatz were used for the guiding functions (GF).
All results are summarized in Table {~\ref{Table1}}.

\vspace{3mm}
In the fine grid calculations, without using the interpolation technique discussed above, a well-converged result was obtained only by using (0.02 Bohr)$^3$ single grid. Indeed, DFT calculation with (0.05 Bohr)$^3$ grid resulted in a much worse DFT-LDA and corresponding VMC-JDFT energies and a very coarse (0.10 Bohr)$^3$ grid implies numerical instabilities. 
On the other hand, a very coarse (0.10 Bohr)$^3$ grid is already sufficient to obtain a reasonable trial wave function when the cubic interpolation method is used, and (0.05 Bohr)$^3$ is essentially converged.
As expected, the DFT energy obtained by the interpolation method (-162.9714 Ha) is not consistent with that obtained by the very fine mesh (-161.4161 Ha) due to the approximation in the Hartree potential.
However, the wave function obtained in this way can be used as a trial wave function for accurate VMC and DMC calculations
because the nodal surface is almost the same as the fine-grid one:
The VMC-JDFT calculations (i.e. only the amplitude is optimized) show that the VMC energy obtained by the interpolation grid (-162.20151(21) Ha) is almost the same as the fine-grid one (-162.20442(21) Ha),
where the deviation is only a few mHartree.
The LRDMC (GF=JDFT) calculations also show a very good agreement (-162.23764(24) Ha and -162.23924(23) Ha for the cubic interpolation and fine grid, respectively).
Furthermore, our LRDMC (GF=JDFT) also reproduces the reference energy (-162.23966(22) Ha) that was obtained by an all-electron DMC (GF=STO-HF) calculation using a very large basis set (quadruple-$\xi$-fourfold-polarized: QZ4P).
These results indicate that the nodal surface determined by the interpolated DFT is as good as the fine-grid and the large-basis one. 
Thus, the interpolation method enables us to obtain a reasonable trial wave function with a low computational cost.
Notice that, this interpolation method was applied also with (0.20 Bohr)$^3$ and (0.01 Bohr)$^3$ double mesh grids with much worse results as far as the quality of the nodal surface and corresponding VMC energies are concerned. Nevertheless , with such a sizable error,  it can be clearly appreciated that the cubic interpolation performs better than the linear one.

\vspace{3mm}
The wave function can be further improved by optimizing the determinant part in presence of the Jastrow factor.
As shown in Table {~\ref{Table1}}, VMC-JSD and VMC-AGP show lower variational energies than VMC-DFT, and LRDMC(GF=JSD, JAGP) also show lower variational energies than LRDMC(GF=JDFT) thanks to the improvement of the nodal surfaces. Remarkably our LRDMC energy corresponding to our best VMC-JAGP is very close to the estimated exact total energy, namely -162.2546 (Ha){~\cite{1993CHA}}.


\section{Application to the sodium dimer}

PESs of the sodium dimer were calculated by VMC-JAGP and LRDMC, by using the developed interpolation scheme, and were compared with previous experiments and calculations. 
First, a PES was calculated using JAGP ansatz by changing the internuclear distance from 1.8 $\AA$ to 10.0 $\AA$, then, a PES was again calculated by LRDMC starting from the optimized wave functions. The energies obtained by LRDMC for each $a$ were extrapolated by quartic polynomial fits $E\left( a \right) = {E_0} + b{a^2} + c{a^4}$ to obtain the $a \to 0$ limit (${E_0}$), wherein $a$ = 0.03, 0.04, 0.05, 0.06, 0.07, and 0.08 are employed (Fig.{~\ref{Figure2}}) in all these calculations. The VMC-JAGP and LRDMC energies are summarized in Table{~\ref{Table2}}, and the obtained PESs are shown in Fig.{~\ref{Figure3}}. The PESs obtained by HF, MP2, CCSD(T)
{\footnote{
The aug-cc-VQZ basis sets were used for these calculations. 
}}
calculated using Gaussian 09, Revision E.01{~\cite{2013FRI}}, and the experimental values reported by Verma et. al {~\cite{1983VER}} are also shown in Fig.{~\ref{Figure3}} for comparison. 
Notice that the energy of the molecule at large distance coincides with twice the energy of a single atom, namely the size consistency is perfectly fulfilled within VMC-JAGP and LRDMC calculations (see the bottom of Table{~\ref{Table2}}).

\vspace{3mm}
We have analyzed the PESs, by using the simple analytic Murrell-Sorbie (MS) function{~\cite{1975SOR}} that has been widely used for describing PES of neutral dimers,
\begin{equation}
E\left( \rho \right) = - {D_\text{e}}\left( {1 + {a_1}\rho + {a_2}{\rho ^2} + {a_3}{\rho ^3}} \right)\exp \left( { - {a_1}\rho } \right),
\end{equation}
where $D_\text{e}$ is the dissociation energy without the zero point vibration energy (ZPVE), $\rho = d - d_\text{eq}$, $d$ is the internuclear distance between the sodium atoms, $d_\text{eq}$ is the equilibrium internuclear distance, and ${a_1}$, ${a_2}$, and ${a_3}$ are fitting parameters. 
The $D_\text{e}$, $d_\text{eq}$, ${a_1}$, ${a_2}$, and ${a_3}$ were determined using scipy.optimize.curve\_fit module implemented in the Python SciPy library{~\cite{2001SCI}}. 
Then, a harmonic vibration frequency (${\omega _\text{e}}$ cm$^{-1}$) was calculated according to the following relation{~\cite{2007BEN}}:
\begin{equation}
{\omega _\text{e}} = \cfrac{1}{{2\pi c}}\sqrt {\cfrac{{{D_\text{e}}\left( {a_1^2 - 2{a_2}} \right)}}{\mu }}, \\
\end{equation}
where $c$ is the light velocity, and $\mu$ is the reduced mass.
The obtained values are summarized in Table{~\ref{Table3}}.

\vspace{3mm}
The VMC calculation reproduces the qualitative shape of the experimental PES (Fig.{~\ref{Figure3}}),
and is accurate for determining the equilibrium distance and the harmonic frequency (Table{~\ref{Table3}}).
However, it is not enough for obtaining the accurate dissociation energy and for reproducing
the binding character in the range of 5.0 $\AA$ and 8.0 $\AA$ 
(i.e. showing higher energies than the dissociation limit).
Notice that the VMC-JAGP is usually enough to 
reproduce almost correct binding energies for the second-row dimers{~{\cite{2009MAR}}.
This suggests that DMC is extremely important for molecules of large atomic number and that our Jastrow is not enough accurate for describing this weak chemical bond.

\vspace{3mm}
The experimental PES is accurately described after the application of the LRDMC projection to the JAGP (Fig.{~\ref{Figure3}}). The equilibrium distance ($d_\text{eq}$ = 3.083(11) \AA) and the harmonic frequency (${\omega _\text{e}}$ = 163.3(3.4) cm$^{-1}$) obtained by our LRDMC calculations are well consistent with the experimental values ($d_\text{eq}$ = 3.08 \AA, ${\omega _\text{e}}$ = 159.1 cm$^{-1}$, and $d_\text{eq}$ = 3.079 \AA, ${\omega _\text{e}}$ = 159.12 cm$^{-1}$ cited from Ref.{~\onlinecite{1995LIU}} and {~\onlinecite{2013HUB}}, respectively).
The dissociation energy ($D_\text{e}$ = 25.28(43) mHa) is also consistent with the experimental ones ($D_\text{e}$ = 26.82 mHa{~\cite{2013HUB}} and $D_\text{e}$ = 27.44 mHa{~\cite{1995LIU}}), and several theoretical works, such as coupled cluster calculations ($D_\text{e}$ = 26.49 mHa for CCSD(T) and $D_\text{e}$ = 26.53 mHa for QCISD{~\cite{2007BEN}}) and full valence configuration interaction calculation ($D_\text{e}$ = 26.85 mHa{~\cite{1993MAG}}), converged within the chemical accuracy ($\sim$ 1kcal/mol $\approx$ 1.6 mHa).

\vspace{3mm}
Although the deviation of the dissociation energy is small enough, it is worth discussing how to obtain a more accurate result. Nemec et. al argued in their work{~\cite{2010NEM}} that the deviation was due to insufficient nodal error cancellation between the atoms and the dimer.
They reported that the dissociation energy of the sodium dimer was underestimated ($D_\text{e}$ = 23.87(57) mHa at $d$ = 3.0789 \AA{\footnote{The original value is 14.918$\pm$0.357 kcal/mol at $d$ = 3.0789 \AA. See their supplementary material.}}) by an all-electron DMC calculation starting from STO{~\cite{2010NEM}}.
The reduction of the error cancellation is important to obtain a more accurate result.
In order to compare directly our results with the previous DMC one{~\cite{2010NEM}}, 
LRDMC (GF=JDFT, JSD, JAGP) for the sodium dimer at the same distance ($d$ = 3.0789 \AA) were performed. The results are shown in Fig.~{\ref{Figure4}} and summarized in Table~{\ref{Table4}}.
LRDMC (GF=JDFT) and LRDMC (GF=JSD) give $D_\text{e}$ = 23.23(60) mHa and $D_\text{e}$ =23.47(50) mHa, respectively, which are statistically consistent with the previous DMC calculation ($D_\text{e}$ = 23.87(57) mHa). 
On the other hand, our LRDMC (GF=JAGP) greatly improves the error cancellation and provides, to our knowledge, the best binding energy ($D_\text{e}$ = 25.34(60) mHa) so far available within QMC techniques.
Fig.~{\ref{Figure4}} shows that the best variational energies are obtained when LRDMC is applied to the JAGP guiding functions both for the atom and the dimer, meaning that the corresponding nodal surfaces are better than previous calculations.
Table.~{\ref{Table5}} summarizes the absolute energies of the sodium atom and dimer, and the residual errors in the absolute energies and corresponding binding energies within the fixed-node approximation.
The nodal surface errors in LRDMC (GF=JDFT) are 33.92 mHa and 37.51 mHa for two sodium atoms and for the dimer, respectively. This implies 3.59 mHa smaller binding energy ($D_\text{e}$ = 23.23(60) mHa) than the experimental value ($D_\text{e}$ = 26.82 mHa) due to insufficient error cancellation.
On the other hand, the nodal surface errors in LRDMC (GF=JAGP) become smaller, 24.21 mHa and 25.69 mHa for two sodium atoms and the dimer, respectively, thanks to the multiconfigurational nature of JAGP{~\cite{2014ZEN, 2019GEN}} (i.e. static correlation). This leads to a much better binding energy ($D_\text{e}$ = 25.34(60) mHa) because of the improvement in the error cancellation. 
Figure~{\ref{Figure5}} shows the energy diagram and the results of the error cancellations.
Compared to LRDMC (GF=JDFT), LRDMC (GF=JAGP) reduces by 9.71 mHa the nodal error for the two sodium atoms, while by 11.82 mHa for the dimer, resulting in a better error cancellation and a corresponding more accurate binding energy. While the value of Ref.{~\onlinecite{2013HUB}} is used for the exact binding energy in this discussion, the conclusion does not change when the other experimental value (e.g. 27.44 mHa{~\cite{1995LIU}}) is employed.
The fact that LRDMC (GF=JAGP) lowers the total energy more effectively in the dimer rather than in the atom indicates that the DFT nodes have some error 
also in the valence region because one can assume an almost exact nodal error cancellation in the core region{~\cite{2010NEM}}.
Thus, we expect that the use of more flexible wave functions such as backflow{~\cite{2006ROP}} or pfaffian{~\cite{2006BAJ}} should further reduce the error and should lead to an almost exact error cancellation (i.e. better binding energy) and essentially exact binding energies of dimers as well as PESs.


\section{Summary}

In this work, we report potential energy surfaces (PES) of the sodium dimer calculated by variational (VMC) and lattice regularized diffusion Monte Carlo (LRDMC). 
Remarkably, after the application of the LRDMC projection to the Jastrow Antisymmetrized Geminal Product (JAGP) ansatz, chemical accuracy is reached, and the obtained dissociation energy, equilibrium internuclear distance, and harmonic vibration frequency are in good agreement with the experimental ones. The trial wave functions for the VMC and LRDMC calculations were prepared by the DFT single determinant ansatz expanded exactly in the same atomic basis used for the QMC calculation, which we have conveniently devised to satisfy exactly the electron-ion cusp conditions. This allows us to use a very small basis, which is however almost converged in the complete basis set limit. In this way, it is possible to reduce the computational effort as well as the statistical fluctuations on the total energy.
We have found that the improvement in the description of the electron correlation and the weak chemical bond of the sodium dimer, is mainly achieved thanks to the energy optimization strategy that we have developed in this work.
For the all-electron calculation, the DFT step is computationally very demanding, at least in the convenient basis we have chosen. 
Therefore, we have developed an efficient DFT algorithm in the mentioned basis, by estimating the corresponding matrix elements on a mesh, and by using a much finer mesh grid only in the vicinity of the nuclei.
In this way, we can have a very good description of this chemical bond and evaluate the corresponding PES with a high degree of accuracy. We believe that our work represents an important step to define a quantum Monte Carlo method that will have the same reliability and accuracy of modern quantum chemistry packages in the future, with the considerable advantage that QMC with the single determinant ansatz used in this work scales very well with the number of electrons and has an almost ideal scaling for massively parallel computations. 


\section{Acknowledgements}
The computations in this work have been mainly performed using the facilities of Research Center for Advanced Computing Infrastructure at Japan Advanced Institute of Science and Technology (JAIST). The structure of the sodium dimer was depicted by VESTA 3 package{~\cite{2011MOM}}.
K. Nakano is grateful for a financial support from Simons Foundation and for technical support by Prof. K. Hongo (JAIST). K. Nakano and S. Sorella are grateful for useful discussion with C. Genovese (SISSA).
R. Maezono is grateful for financial supports from MEXT-KAKENHI (17H05478 and 16KK0097), from FLAGSHIP2020 (project nos. hp180206 and hp180175 at K-computer), from Toyota Motor Corporation, from I-O DATA Foundation, and from the Air Force Office of Scientific Research (AFOSR-AOARD/FA2386-17-1-4049).

\bibliographystyle{apsrev4-1}
\bibliography{./references.bib}

\begin{thebibliography}{88}%
\makeatletter
\providecommand \@ifxundefined [1]{%
 \@ifx{#1\undefined}
}%
\providecommand \@ifnum [1]{%
 \ifnum #1\expandafter \@firstoftwo
 \else \expandafter \@secondoftwo
 \fi
}%
\providecommand \@ifx [1]{%
 \ifx #1\expandafter \@firstoftwo
 \else \expandafter \@secondoftwo
 \fi
}%
\providecommand \natexlab [1]{#1}%
\providecommand \enquote  [1]{``#1''}%
\providecommand \bibnamefont  [1]{#1}%
\providecommand \bibfnamefont [1]{#1}%
\providecommand \citenamefont [1]{#1}%
\providecommand \href@noop [0]{\@secondoftwo}%
\providecommand \href [0]{\begingroup \@sanitize@url \@href}%
\providecommand \@href[1]{\@@startlink{#1}\@@href}%
\providecommand \@@href[1]{\endgroup#1\@@endlink}%
\providecommand \@sanitize@url [0]{\catcode `\\12\catcode `\$12\catcode
  `\&12\catcode `\#12\catcode `\^12\catcode `\_12\catcode `\%12\relax}%
\providecommand \@@startlink[1]{}%
\providecommand \@@endlink[0]{}%
\providecommand \url  [0]{\begingroup\@sanitize@url \@url }%
\providecommand \@url [1]{\endgroup\@href {#1}{\urlprefix }}%
\providecommand \urlprefix  [0]{URL }%
\providecommand \Eprint [0]{\href }%
\providecommand \doibase [0]{http://dx.doi.org/}%
\providecommand \selectlanguage [0]{\@gobble}%
\providecommand \bibinfo  [0]{\@secondoftwo}%
\providecommand \bibfield  [0]{\@secondoftwo}%
\providecommand \translation [1]{[#1]}%
\providecommand \BibitemOpen [0]{}%
\providecommand \bibitemStop [0]{}%
\providecommand \bibitemNoStop [0]{.\EOS\space}%
\providecommand \EOS [0]{\spacefactor3000\relax}%
\providecommand \BibitemShut  [1]{\csname bibitem#1\endcsname}%
\let\auto@bib@innerbib\@empty
\bibitem [{\citenamefont {Foulkes}\ \emph {et~al.}(2001)\citenamefont
  {Foulkes}, \citenamefont {Mitas}, \citenamefont {Needs},\ and\ \citenamefont
  {Rajagopal}}]{2001FOU}%
  \BibitemOpen
  \bibfield  {author} {\bibinfo {author} {\bibfnamefont {W.}~\bibnamefont
  {Foulkes}}, \bibinfo {author} {\bibfnamefont {L.}~\bibnamefont {Mitas}},
  \bibinfo {author} {\bibfnamefont {R.}~\bibnamefont {Needs}}, \ and\ \bibinfo
  {author} {\bibfnamefont {G.}~\bibnamefont {Rajagopal}},\ }\href@noop {}
  {\bibfield  {journal} {\bibinfo  {journal} {Rev. Mod. Phys.}\ }\textbf
  {\bibinfo {volume} {73}},\ \bibinfo {pages} {33} (\bibinfo {year}
  {2001})}\BibitemShut {NoStop}%
\bibitem [{\citenamefont {Luo}\ \emph {et~al.}(2014)\citenamefont {Luo},
  \citenamefont {Zen},\ and\ \citenamefont {Sorella}}]{2014LUO}%
  \BibitemOpen
  \bibfield  {author} {\bibinfo {author} {\bibfnamefont {Y.}~\bibnamefont
  {Luo}}, \bibinfo {author} {\bibfnamefont {A.}~\bibnamefont {Zen}}, \ and\
  \bibinfo {author} {\bibfnamefont {S.}~\bibnamefont {Sorella}},\ }\href@noop
  {} {\bibfield  {journal} {\bibinfo  {journal} {J. Chem. Phys.}\ }\textbf
  {\bibinfo {volume} {141}},\ \bibinfo {pages} {194112} (\bibinfo {year}
  {2014})}\BibitemShut {NoStop}%
\bibitem [{\citenamefont {Zen}\ \emph {et~al.}(2013)\citenamefont {Zen},
  \citenamefont {Luo}, \citenamefont {Sorella},\ and\ \citenamefont
  {Guidoni}}]{2013ZEN}%
  \BibitemOpen
  \bibfield  {author} {\bibinfo {author} {\bibfnamefont {A.}~\bibnamefont
  {Zen}}, \bibinfo {author} {\bibfnamefont {Y.}~\bibnamefont {Luo}}, \bibinfo
  {author} {\bibfnamefont {S.}~\bibnamefont {Sorella}}, \ and\ \bibinfo
  {author} {\bibfnamefont {L.}~\bibnamefont {Guidoni}},\ }\href@noop {}
  {\bibfield  {journal} {\bibinfo  {journal} {J. Chem. Theory Comput.}\
  }\textbf {\bibinfo {volume} {9}},\ \bibinfo {pages} {4332} (\bibinfo {year}
  {2013})}\BibitemShut {NoStop}%
\bibitem [{\citenamefont {Mussard}\ \emph {et~al.}(2018)\citenamefont
  {Mussard}, \citenamefont {Coccia}, \citenamefont {Assaraf}, \citenamefont
  {Otten}, \citenamefont {Umrigar},\ and\ \citenamefont {Toulouse}}]{2018MUS}%
  \BibitemOpen
  \bibfield  {author} {\bibinfo {author} {\bibfnamefont {B.}~\bibnamefont
  {Mussard}}, \bibinfo {author} {\bibfnamefont {E.}~\bibnamefont {Coccia}},
  \bibinfo {author} {\bibfnamefont {R.}~\bibnamefont {Assaraf}}, \bibinfo
  {author} {\bibfnamefont {M.}~\bibnamefont {Otten}}, \bibinfo {author}
  {\bibfnamefont {C.~J.}\ \bibnamefont {Umrigar}}, \ and\ \bibinfo {author}
  {\bibfnamefont {J.}~\bibnamefont {Toulouse}},\ }in\ \href@noop {} {\emph
  {\bibinfo {booktitle} {Advances in Quantum Chemistry}}},\ Vol.~\bibinfo
  {volume} {76}\ (\bibinfo  {publisher} {Academic Press},\ \bibinfo {year}
  {2018})\ pp.\ \bibinfo {pages} {255--270}\BibitemShut {NoStop}%
\bibitem [{\citenamefont {Hunt}\ \emph {et~al.}(2018)\citenamefont {Hunt},
  \citenamefont {Szyniszewski}, \citenamefont {Prayogo}, \citenamefont
  {Maezono},\ and\ \citenamefont {Drummond}}]{2018HUN}%
  \BibitemOpen
  \bibfield  {author} {\bibinfo {author} {\bibfnamefont {R.~J.}\ \bibnamefont
  {Hunt}}, \bibinfo {author} {\bibfnamefont {M.}~\bibnamefont {Szyniszewski}},
  \bibinfo {author} {\bibfnamefont {G.~I.}\ \bibnamefont {Prayogo}}, \bibinfo
  {author} {\bibfnamefont {R.}~\bibnamefont {Maezono}}, \ and\ \bibinfo
  {author} {\bibfnamefont {N.~D.}\ \bibnamefont {Drummond}},\ }\href {\doibase
  10.1103/PhysRevB.98.075122} {\bibfield  {journal} {\bibinfo  {journal} {Phys.
  Rev. B}\ }\textbf {\bibinfo {volume} {98}},\ \bibinfo {pages} {075122}
  (\bibinfo {year} {2018})}\BibitemShut {NoStop}%
\bibitem [{\citenamefont {Mazzola}\ \emph {et~al.}(2018)\citenamefont
  {Mazzola}, \citenamefont {Helled},\ and\ \citenamefont {Sorella}}]{2018MAZ}%
  \BibitemOpen
  \bibfield  {author} {\bibinfo {author} {\bibfnamefont {G.}~\bibnamefont
  {Mazzola}}, \bibinfo {author} {\bibfnamefont {R.}~\bibnamefont {Helled}}, \
  and\ \bibinfo {author} {\bibfnamefont {S.}~\bibnamefont {Sorella}},\
  }\href@noop {} {\bibfield  {journal} {\bibinfo  {journal} {Phys. Rev. Lett.}\
  }\textbf {\bibinfo {volume} {120}},\ \bibinfo {pages} {025701} (\bibinfo
  {year} {2018})}\BibitemShut {NoStop}%
\bibitem [{\citenamefont {Stella}\ \emph {et~al.}(2011)\citenamefont {Stella},
  \citenamefont {Attaccalite}, \citenamefont {Sorella},\ and\ \citenamefont
  {Rubio}}]{2011STE}%
  \BibitemOpen
  \bibfield  {author} {\bibinfo {author} {\bibfnamefont {L.}~\bibnamefont
  {Stella}}, \bibinfo {author} {\bibfnamefont {C.}~\bibnamefont {Attaccalite}},
  \bibinfo {author} {\bibfnamefont {S.}~\bibnamefont {Sorella}}, \ and\
  \bibinfo {author} {\bibfnamefont {A.}~\bibnamefont {Rubio}},\ }\href@noop {}
  {\bibfield  {journal} {\bibinfo  {journal} {Phys. Rev. B}\ }\textbf {\bibinfo
  {volume} {84}},\ \bibinfo {pages} {245117} (\bibinfo {year}
  {2011})}\BibitemShut {NoStop}%
\bibitem [{\citenamefont {Varsano}\ \emph {et~al.}(2017)\citenamefont
  {Varsano}, \citenamefont {Sorella}, \citenamefont {Sangalli}, \citenamefont
  {Barborini}, \citenamefont {Corni}, \citenamefont {Molinari},\ and\
  \citenamefont {Rontani}}]{2017VAR}%
  \BibitemOpen
  \bibfield  {author} {\bibinfo {author} {\bibfnamefont {D.}~\bibnamefont
  {Varsano}}, \bibinfo {author} {\bibfnamefont {S.}~\bibnamefont {Sorella}},
  \bibinfo {author} {\bibfnamefont {D.}~\bibnamefont {Sangalli}}, \bibinfo
  {author} {\bibfnamefont {M.}~\bibnamefont {Barborini}}, \bibinfo {author}
  {\bibfnamefont {S.}~\bibnamefont {Corni}}, \bibinfo {author} {\bibfnamefont
  {E.}~\bibnamefont {Molinari}}, \ and\ \bibinfo {author} {\bibfnamefont
  {M.}~\bibnamefont {Rontani}},\ }\href {\doibase 10.1038/s41467-017-01660-8}
  {\bibfield  {journal} {\bibinfo  {journal} {Nat. Commun.}\ }\textbf {\bibinfo
  {volume} {8}},\ \bibinfo {pages} {1461} (\bibinfo {year} {2017})}\BibitemShut
  {NoStop}%
\bibitem [{\citenamefont {Ceperley}(1986)}]{1986CEP}%
  \BibitemOpen
  \bibfield  {author} {\bibinfo {author} {\bibfnamefont {D.}~\bibnamefont
  {Ceperley}},\ }\href@noop {} {\bibfield  {journal} {\bibinfo  {journal} {J.
  Stat. Phys.}\ }\textbf {\bibinfo {volume} {43}},\ \bibinfo {pages} {815}
  (\bibinfo {year} {1986})}\BibitemShut {NoStop}%
\bibitem [{\citenamefont {Hammond}\ \emph {et~al.}(1987)\citenamefont
  {Hammond}, \citenamefont {Reynolds},\ and\ \citenamefont
  {Lester~Jr}}]{1987HAM}%
  \BibitemOpen
  \bibfield  {author} {\bibinfo {author} {\bibfnamefont {B.~L.}\ \bibnamefont
  {Hammond}}, \bibinfo {author} {\bibfnamefont {P.~J.}\ \bibnamefont
  {Reynolds}}, \ and\ \bibinfo {author} {\bibfnamefont {W.~A.}\ \bibnamefont
  {Lester~Jr}},\ }\href@noop {} {\bibfield  {journal} {\bibinfo  {journal} {J.
  Chem. Phys.}\ }\textbf {\bibinfo {volume} {87}},\ \bibinfo {pages} {1130}
  (\bibinfo {year} {1987})}\BibitemShut {NoStop}%
\bibitem [{\citenamefont {Ma}\ \emph {et~al.}(2005)\citenamefont {Ma},
  \citenamefont {Drummond}, \citenamefont {Towler},\ and\ \citenamefont
  {Needs}}]{2005MA}%
  \BibitemOpen
  \bibfield  {author} {\bibinfo {author} {\bibfnamefont {A.}~\bibnamefont
  {Ma}}, \bibinfo {author} {\bibfnamefont {N.}~\bibnamefont {Drummond}},
  \bibinfo {author} {\bibfnamefont {M.}~\bibnamefont {Towler}}, \ and\ \bibinfo
  {author} {\bibfnamefont {R.}~\bibnamefont {Needs}},\ }\href@noop {}
  {\bibfield  {journal} {\bibinfo  {journal} {Phys. Rev. E}\ }\textbf {\bibinfo
  {volume} {71}},\ \bibinfo {pages} {066704} (\bibinfo {year}
  {2005})}\BibitemShut {NoStop}%
\bibitem [{\citenamefont {Trail}\ and\ \citenamefont {Needs}(2015)}]{2015TRA}%
  \BibitemOpen
  \bibfield  {author} {\bibinfo {author} {\bibfnamefont {J.~R.}\ \bibnamefont
  {Trail}}\ and\ \bibinfo {author} {\bibfnamefont {R.~J.}\ \bibnamefont
  {Needs}},\ }\href {\doibase 10.1063/1.4907589} {\bibfield  {journal}
  {\bibinfo  {journal} {J. Chem. Phys.}\ }\textbf {\bibinfo {volume} {142}},\
  \bibinfo {pages} {064110} (\bibinfo {year} {2015})}\BibitemShut {NoStop}%
\bibitem [{\citenamefont {Krogel}\ \emph {et~al.}(2016)\citenamefont {Krogel},
  \citenamefont {Santana},\ and\ \citenamefont {Reboredo}}]{2016KRO}%
  \BibitemOpen
  \bibfield  {author} {\bibinfo {author} {\bibfnamefont {J.~T.}\ \bibnamefont
  {Krogel}}, \bibinfo {author} {\bibfnamefont {J.~A.}\ \bibnamefont {Santana}},
  \ and\ \bibinfo {author} {\bibfnamefont {F.~A.}\ \bibnamefont {Reboredo}},\
  }\href {\doibase 10.1103/PhysRevB.93.075143} {\bibfield  {journal} {\bibinfo
  {journal} {Phys. Rev. B}\ }\textbf {\bibinfo {volume} {93}},\ \bibinfo
  {pages} {75143} (\bibinfo {year} {2016})}\BibitemShut {NoStop}%
\bibitem [{\citenamefont {Trail}\ and\ \citenamefont {Needs}(2017)}]{2017TRA}%
  \BibitemOpen
  \bibfield  {author} {\bibinfo {author} {\bibfnamefont {J.~R.}\ \bibnamefont
  {Trail}}\ and\ \bibinfo {author} {\bibfnamefont {R.~J.}\ \bibnamefont
  {Needs}},\ }\href {\doibase 10.1063/1.4984046} {\bibfield  {journal}
  {\bibinfo  {journal} {J. Chem. Phys.}\ }\textbf {\bibinfo {volume} {146}},\
  \bibinfo {pages} {204107} (\bibinfo {year} {2017})}\BibitemShut {NoStop}%
\bibitem [{\citenamefont {Bennett}\ \emph {et~al.}(2017)\citenamefont
  {Bennett}, \citenamefont {Melton}, \citenamefont {Annaberdiyev},
  \citenamefont {Wang}, \citenamefont {Shulenburger},\ and\ \citenamefont
  {Mitas}}]{2017BEN}%
  \BibitemOpen
  \bibfield  {author} {\bibinfo {author} {\bibfnamefont {M.~C.}\ \bibnamefont
  {Bennett}}, \bibinfo {author} {\bibfnamefont {C.~A.}\ \bibnamefont {Melton}},
  \bibinfo {author} {\bibfnamefont {A.}~\bibnamefont {Annaberdiyev}}, \bibinfo
  {author} {\bibfnamefont {G.}~\bibnamefont {Wang}}, \bibinfo {author}
  {\bibfnamefont {L.}~\bibnamefont {Shulenburger}}, \ and\ \bibinfo {author}
  {\bibfnamefont {L.}~\bibnamefont {Mitas}},\ }\href {\doibase
  10.1063/1.4995643} {\bibfield  {journal} {\bibinfo  {journal} {J. Chem.
  Phys.}\ }\textbf {\bibinfo {volume} {147}},\ \bibinfo {pages} {224106}
  (\bibinfo {year} {2017})}\BibitemShut {NoStop}%
\bibitem [{\citenamefont {Bennett}\ \emph {et~al.}(2018)\citenamefont
  {Bennett}, \citenamefont {Wang}, \citenamefont {Annaberdiyev}, \citenamefont
  {Melton}, \citenamefont {Shulenburger},\ and\ \citenamefont
  {Mitas}}]{2018BEN}%
  \BibitemOpen
  \bibfield  {author} {\bibinfo {author} {\bibfnamefont {M.~C.}\ \bibnamefont
  {Bennett}}, \bibinfo {author} {\bibfnamefont {G.}~\bibnamefont {Wang}},
  \bibinfo {author} {\bibfnamefont {A.}~\bibnamefont {Annaberdiyev}}, \bibinfo
  {author} {\bibfnamefont {C.~A.}\ \bibnamefont {Melton}}, \bibinfo {author}
  {\bibfnamefont {L.}~\bibnamefont {Shulenburger}}, \ and\ \bibinfo {author}
  {\bibfnamefont {L.}~\bibnamefont {Mitas}},\ }\href {\doibase
  10.1063/1.5038135} {\bibfield  {journal} {\bibinfo  {journal} {J. Chem.
  Phys.}\ }\textbf {\bibinfo {volume} {149}},\ \bibinfo {pages} {104108}
  (\bibinfo {year} {2018})}\BibitemShut {NoStop}%
\bibitem [{\citenamefont {Annaberdiyev}\ \emph {et~al.}(2018)\citenamefont
  {Annaberdiyev}, \citenamefont {Wang}, \citenamefont {Melton}, \citenamefont
  {{Chandler Bennett}}, \citenamefont {Shulenburger},\ and\ \citenamefont
  {Mitas}}]{2018ANN}%
  \BibitemOpen
  \bibfield  {author} {\bibinfo {author} {\bibfnamefont {A.}~\bibnamefont
  {Annaberdiyev}}, \bibinfo {author} {\bibfnamefont {G.}~\bibnamefont {Wang}},
  \bibinfo {author} {\bibfnamefont {C.~A.}\ \bibnamefont {Melton}}, \bibinfo
  {author} {\bibfnamefont {M.}~\bibnamefont {{Chandler Bennett}}}, \bibinfo
  {author} {\bibfnamefont {L.}~\bibnamefont {Shulenburger}}, \ and\ \bibinfo
  {author} {\bibfnamefont {L.}~\bibnamefont {Mitas}},\ }\href {\doibase
  10.1063/1.5040472} {\bibfield  {journal} {\bibinfo  {journal} {J. Chem.
  Phys.}\ }\textbf {\bibinfo {volume} {149}},\ \bibinfo {pages} {134108}
  (\bibinfo {year} {2018})}\BibitemShut {NoStop}%
\bibitem [{\citenamefont {Umrigar}(1993)}]{1993UMR}%
  \BibitemOpen
  \bibfield  {author} {\bibinfo {author} {\bibfnamefont {C.}~\bibnamefont
  {Umrigar}},\ }\href@noop {} {\bibfield  {journal} {\bibinfo  {journal} {Phys.
  Rev. Lett.}\ }\textbf {\bibinfo {volume} {71}},\ \bibinfo {pages} {408}
  (\bibinfo {year} {1993})}\BibitemShut {NoStop}%
\bibitem [{\citenamefont {Stedman}\ \emph {et~al.}(1998)\citenamefont
  {Stedman}, \citenamefont {Foulkes},\ and\ \citenamefont {Nekovee}}]{1998STE}%
  \BibitemOpen
  \bibfield  {author} {\bibinfo {author} {\bibfnamefont {M.}~\bibnamefont
  {Stedman}}, \bibinfo {author} {\bibfnamefont {W.}~\bibnamefont {Foulkes}}, \
  and\ \bibinfo {author} {\bibfnamefont {M.}~\bibnamefont {Nekovee}},\
  }\href@noop {} {\bibfield  {journal} {\bibinfo  {journal} {J. Chem. Phys.}\
  }\textbf {\bibinfo {volume} {109}},\ \bibinfo {pages} {2630} (\bibinfo {year}
  {1998})}\BibitemShut {NoStop}%
\bibitem [{\citenamefont {Martin}(2004)}]{2004MAR}%
  \BibitemOpen
  \bibfield  {author} {\bibinfo {author} {\bibfnamefont {R.~M.}\ \bibnamefont
  {Martin}},\ }\href
  {https://books.google.it/books?hl=en&lr=&id=v1YhAwAAQBAJ&oi=fnd&pg=PR17&dq=martin+electronic+structure&ots=8Py35p2fcf&sig=m6qYd3VspZjCrjnB0olxmT4z3zQ#v=onepage&q=martin
  electronic structure&f=false} {\emph {\bibinfo {title} {{Electronic structure
  : basic theory and practical methods}}}}\ (\bibinfo  {publisher} {Cambridge
  University Press},\ \bibinfo {year} {2004})\BibitemShut {NoStop}%
\bibitem [{\citenamefont {Szabo}\ and\ \citenamefont
  {Ostlund}(1982)}]{1996SZA}%
  \BibitemOpen
  \bibfield  {author} {\bibinfo {author} {\bibfnamefont {A.}~\bibnamefont
  {Szabo}}\ and\ \bibinfo {author} {\bibfnamefont {N.}~\bibnamefont
  {Ostlund}},\ }\href@noop {} {\emph {\bibinfo {title} {{Modern quantum
  chemistry}}}}\ (\bibinfo  {publisher} {Macmillan PubliShing Co., Inc.},\
  \bibinfo {year} {1982})\BibitemShut {NoStop}%
\bibitem [{\citenamefont {Sorella}()}]{2019SOR}%
  \BibitemOpen
  \bibfield  {author} {\bibinfo {author} {\bibfnamefont {S.}~\bibnamefont
  {Sorella}},\ }\href@noop {} {\enquote {\bibinfo {title} {{TurboRVB:Quantum
  Monte Carlo Software for Electronic Structure Calculations}},}\ }\bibinfo
  {howpublished} {\url{https://people.sissa.it/~sorella/web}},\ \bibinfo {note}
  {[Online; accessed 10-March-2019]}\BibitemShut {NoStop}%
\bibitem [{\citenamefont {Beck}(2000)}]{2000BEC}%
  \BibitemOpen
  \bibfield  {author} {\bibinfo {author} {\bibfnamefont {T.~L.}\ \bibnamefont
  {Beck}},\ }\href {\doibase 10.1103/RevModPhys.72.1041} {\bibfield  {journal}
  {\bibinfo  {journal} {Rev. Mod. Phys.}\ }\textbf {\bibinfo {volume} {72}},\
  \bibinfo {pages} {1041} (\bibinfo {year} {2000})}\BibitemShut {NoStop}%
\bibitem [{\citenamefont {Singh}(1994)}]{1994SIN}%
  \BibitemOpen
  \bibfield  {author} {\bibinfo {author} {\bibfnamefont {D.~J.}\ \bibnamefont
  {Singh}},\ }\href {\doibase 10.1007/978-1-4757-2312-0} {\emph {\bibinfo
  {title} {{Planewaves, Pseudopotentials and the LAPW Method}}}}\ (\bibinfo
  {publisher} {Springer US},\ \bibinfo {address} {Boston, MA},\ \bibinfo {year}
  {1994})\BibitemShut {NoStop}%
\bibitem [{\citenamefont {Wood}\ and\ \citenamefont {Hackett}(1909)}]{1909WOO}%
  \BibitemOpen
  \bibfield  {author} {\bibinfo {author} {\bibfnamefont {R.~W.}\ \bibnamefont
  {Wood}}\ and\ \bibinfo {author} {\bibfnamefont {F.~E.}\ \bibnamefont
  {Hackett}},\ }\href@noop {} {\bibfield  {journal} {\bibinfo  {journal}
  {Astrophys. J.}\ }\textbf {\bibinfo {volume} {30}},\ \bibinfo {pages} {339}
  (\bibinfo {year} {1909})}\BibitemShut {NoStop}%
\bibitem [{\citenamefont {Fredrickson}(1929)}]{1929FRE}%
  \BibitemOpen
  \bibfield  {author} {\bibinfo {author} {\bibfnamefont {W.~R.}\ \bibnamefont
  {Fredrickson}},\ }\href@noop {} {\bibfield  {journal} {\bibinfo  {journal}
  {Phys. Rev.}\ }\textbf {\bibinfo {volume} {34}},\ \bibinfo {pages} {207}
  (\bibinfo {year} {1929})}\BibitemShut {NoStop}%
\bibitem [{\citenamefont {Demtr{\"{o}}der}\ \emph {et~al.}(1969)\citenamefont
  {Demtr{\"{o}}der}, \citenamefont {McClintock},\ and\ \citenamefont
  {Zare}}]{1969DEM}%
  \BibitemOpen
  \bibfield  {author} {\bibinfo {author} {\bibfnamefont {W.}~\bibnamefont
  {Demtr{\"{o}}der}}, \bibinfo {author} {\bibfnamefont {M.}~\bibnamefont
  {McClintock}}, \ and\ \bibinfo {author} {\bibfnamefont {R.~N.}\ \bibnamefont
  {Zare}},\ }\href@noop {} {\bibfield  {journal} {\bibinfo  {journal} {J. Chem.
  Phys.}\ }\textbf {\bibinfo {volume} {51}},\ \bibinfo {pages} {5495} (\bibinfo
  {year} {1969})}\BibitemShut {NoStop}%
\bibitem [{\citenamefont {Demtr{\"{o}}der}\ and\ \citenamefont
  {Stock}(1975)}]{1975DEM}%
  \BibitemOpen
  \bibfield  {author} {\bibinfo {author} {\bibfnamefont {W.}~\bibnamefont
  {Demtr{\"{o}}der}}\ and\ \bibinfo {author} {\bibfnamefont {M.}~\bibnamefont
  {Stock}},\ }\href@noop {} {\bibfield  {journal} {\bibinfo  {journal} {J. Mol.
  Spectrosc.}\ }\textbf {\bibinfo {volume} {55}},\ \bibinfo {pages} {476}
  (\bibinfo {year} {1975})}\BibitemShut {NoStop}%
\bibitem [{\citenamefont {Kusch}\ and\ \citenamefont {Hessel}(1978)}]{1978KUS}%
  \BibitemOpen
  \bibfield  {author} {\bibinfo {author} {\bibfnamefont {P.}~\bibnamefont
  {Kusch}}\ and\ \bibinfo {author} {\bibfnamefont {M.~M.}\ \bibnamefont
  {Hessel}},\ }\href {\doibase 10.1063/1.436117} {\bibfield  {journal}
  {\bibinfo  {journal} {J. Chem. Phys.}\ }\textbf {\bibinfo {volume} {68}},\
  \bibinfo {pages} {2591} (\bibinfo {year} {1978})}\BibitemShut {NoStop}%
\bibitem [{\citenamefont {Verma}\ \emph {et~al.}(1983)\citenamefont {Verma},
  \citenamefont {Bahns}, \citenamefont {Rajaei-Rizi}, \citenamefont
  {Stwalley},\ and\ \citenamefont {Zemke}}]{1983VER}%
  \BibitemOpen
  \bibfield  {author} {\bibinfo {author} {\bibfnamefont {K.}~\bibnamefont
  {Verma}}, \bibinfo {author} {\bibfnamefont {J.}~\bibnamefont {Bahns}},
  \bibinfo {author} {\bibfnamefont {A.}~\bibnamefont {Rajaei-Rizi}}, \bibinfo
  {author} {\bibfnamefont {W.~C.}\ \bibnamefont {Stwalley}}, \ and\ \bibinfo
  {author} {\bibfnamefont {W.}~\bibnamefont {Zemke}},\ }\href@noop {}
  {\bibfield  {journal} {\bibinfo  {journal} {J. Chem. Phys.}\ }\textbf
  {\bibinfo {volume} {78}},\ \bibinfo {pages} {3599} (\bibinfo {year}
  {1983})}\BibitemShut {NoStop}%
\bibitem [{\citenamefont {Kaminsky}(1977)}]{1997KAM}%
  \BibitemOpen
  \bibfield  {author} {\bibinfo {author} {\bibfnamefont {M.~E.}\ \bibnamefont
  {Kaminsky}},\ }\href {\doibase 10.1063/1.433793} {\bibfield  {journal}
  {\bibinfo  {journal} {J. Chem. Phys.}\ }\textbf {\bibinfo {volume} {66}},\
  \bibinfo {pages} {4951} (\bibinfo {year} {1977})}\BibitemShut {NoStop}%
\bibitem [{\citenamefont {Qi}\ \emph {et~al.}(2007)\citenamefont {Qi},
  \citenamefont {Bai}, \citenamefont {Ahmed}, \citenamefont {Lyyra},
  \citenamefont {Kotochigova}, \citenamefont {Ross}, \citenamefont {Effantin},
  \citenamefont {Zalicki}, \citenamefont {Vig{\'{u}}}, \citenamefont {Chawla},
  \citenamefont {Field}, \citenamefont {Whang}, \citenamefont {Stwalley},
  \citenamefont {Kn{\"{o}}ckel}, \citenamefont {Tiemann}, \citenamefont
  {Shang}, \citenamefont {Li},\ and\ \citenamefont {Bergeman}}]{2007BAI}%
  \BibitemOpen
  \bibfield  {author} {\bibinfo {author} {\bibfnamefont {P.}~\bibnamefont
  {Qi}}, \bibinfo {author} {\bibfnamefont {J.}~\bibnamefont {Bai}}, \bibinfo
  {author} {\bibfnamefont {E.}~\bibnamefont {Ahmed}}, \bibinfo {author}
  {\bibfnamefont {A.~M.}\ \bibnamefont {Lyyra}}, \bibinfo {author}
  {\bibfnamefont {S.}~\bibnamefont {Kotochigova}}, \bibinfo {author}
  {\bibfnamefont {A.~J.}\ \bibnamefont {Ross}}, \bibinfo {author}
  {\bibfnamefont {C.}~\bibnamefont {Effantin}}, \bibinfo {author}
  {\bibfnamefont {P.}~\bibnamefont {Zalicki}}, \bibinfo {author} {\bibfnamefont
  {J.}~\bibnamefont {Vig{\'{u}}}}, \bibinfo {author} {\bibfnamefont
  {G.}~\bibnamefont {Chawla}}, \bibinfo {author} {\bibfnamefont {R.~W.}\
  \bibnamefont {Field}}, \bibinfo {author} {\bibfnamefont {T.~J.}\ \bibnamefont
  {Whang}}, \bibinfo {author} {\bibfnamefont {W.~C.}\ \bibnamefont {Stwalley}},
  \bibinfo {author} {\bibfnamefont {H.}~\bibnamefont {Kn{\"{o}}ckel}}, \bibinfo
  {author} {\bibfnamefont {E.}~\bibnamefont {Tiemann}}, \bibinfo {author}
  {\bibfnamefont {J.}~\bibnamefont {Shang}}, \bibinfo {author} {\bibfnamefont
  {L.}~\bibnamefont {Li}}, \ and\ \bibinfo {author} {\bibfnamefont
  {T.}~\bibnamefont {Bergeman}},\ }\href {\doibase 10.1063/1.2747595}
  {\bibfield  {journal} {\bibinfo  {journal} {J. Chem. Phys.}\ }\textbf
  {\bibinfo {volume} {127}},\ \bibinfo {pages} {044301} (\bibinfo {year}
  {2007})}\BibitemShut {NoStop}%
\bibitem [{\citenamefont {Magnier}\ \emph {et~al.}(1993)\citenamefont
  {Magnier}, \citenamefont {Milli{\'e}}, \citenamefont {Dulieu},\ and\
  \citenamefont {Masnou-Seeuws}}]{1993MAG}%
  \BibitemOpen
  \bibfield  {author} {\bibinfo {author} {\bibfnamefont {S.}~\bibnamefont
  {Magnier}}, \bibinfo {author} {\bibfnamefont {P.}~\bibnamefont {Milli{\'e}}},
  \bibinfo {author} {\bibfnamefont {O.}~\bibnamefont {Dulieu}}, \ and\ \bibinfo
  {author} {\bibfnamefont {F.}~\bibnamefont {Masnou-Seeuws}},\ }\href@noop {}
  {\bibfield  {journal} {\bibinfo  {journal} {J. Chem. Phys.}\ }\textbf
  {\bibinfo {volume} {98}},\ \bibinfo {pages} {7113} (\bibinfo {year}
  {1993})}\BibitemShut {NoStop}%
\bibitem [{\citenamefont {Ho}\ \emph {et~al.}(2000)\citenamefont {Ho},
  \citenamefont {Rabitz},\ and\ \citenamefont {Scoles}}]{2000HO}%
  \BibitemOpen
  \bibfield  {author} {\bibinfo {author} {\bibfnamefont {T.~S.}\ \bibnamefont
  {Ho}}, \bibinfo {author} {\bibfnamefont {H.}~\bibnamefont {Rabitz}}, \ and\
  \bibinfo {author} {\bibfnamefont {G.}~\bibnamefont {Scoles}},\ }\href
  {\doibase 10.1063/1.481269} {\bibfield  {journal} {\bibinfo  {journal} {J.
  Chem. Phys.}\ }\textbf {\bibinfo {volume} {112}},\ \bibinfo {pages} {6218}
  (\bibinfo {year} {2000})}\BibitemShut {NoStop}%
\bibitem [{\citenamefont {Matsunaga}\ and\ \citenamefont
  {Zavitsas}(2004)}]{2004MAT}%
  \BibitemOpen
  \bibfield  {author} {\bibinfo {author} {\bibfnamefont {N.}~\bibnamefont
  {Matsunaga}}\ and\ \bibinfo {author} {\bibfnamefont {A.~A.}\ \bibnamefont
  {Zavitsas}},\ }\href {\doibase 10.1063/1.1648637} {\bibfield  {journal}
  {\bibinfo  {journal} {J. Chem. Phys.}\ }\textbf {\bibinfo {volume} {120}},\
  \bibinfo {pages} {5624} (\bibinfo {year} {2004})}\BibitemShut {NoStop}%
\bibitem [{\citenamefont {Maroulis}(2004)}]{2004MARO}%
  \BibitemOpen
  \bibfield  {author} {\bibinfo {author} {\bibfnamefont {G.}~\bibnamefont
  {Maroulis}},\ }\href {\doibase 10.1063/1.1812737} {\bibfield  {journal}
  {\bibinfo  {journal} {J. Chem. Phys.}\ }\textbf {\bibinfo {volume} {121}},\
  \bibinfo {pages} {10519} (\bibinfo {year} {2004})}\BibitemShut {NoStop}%
\bibitem [{\citenamefont {Harrison}\ and\ \citenamefont
  {Lawson}(2005)}]{2005HAR}%
  \BibitemOpen
  \bibfield  {author} {\bibinfo {author} {\bibfnamefont {J.~F.}\ \bibnamefont
  {Harrison}}\ and\ \bibinfo {author} {\bibfnamefont {D.~B.}\ \bibnamefont
  {Lawson}},\ }\href {\doibase 10.1002/qua.20400} {\bibfield  {journal}
  {\bibinfo  {journal} {Int. J. Quantum Chem.}\ }\textbf {\bibinfo {volume}
  {102}},\ \bibinfo {pages} {1087} (\bibinfo {year} {2005})}\BibitemShut
  {NoStop}%
\bibitem [{\citenamefont {Ben-Hai}\ \emph {et~al.}(2007)\citenamefont
  {Ben-Hai}, \citenamefont {Qi-Run}, \citenamefont {De-Heng},\ and\
  \citenamefont {Yu-Fang}}]{2007BEN}%
  \BibitemOpen
  \bibfield  {author} {\bibinfo {author} {\bibfnamefont {Y.}~\bibnamefont
  {Ben-Hai}}, \bibinfo {author} {\bibfnamefont {D.}~\bibnamefont {Qi-Run}},
  \bibinfo {author} {\bibfnamefont {S.}~\bibnamefont {De-Heng}}, \ and\
  \bibinfo {author} {\bibfnamefont {L.}~\bibnamefont {Yu-Fang}},\ }\href@noop
  {} {\bibfield  {journal} {\bibinfo  {journal} {Chin. Phys.}\ }\textbf
  {\bibinfo {volume} {16}},\ \bibinfo {pages} {2962} (\bibinfo {year}
  {2007})}\BibitemShut {NoStop}%
\bibitem [{\citenamefont {Musia{\l}}\ and\ \citenamefont
  {Bartlett}(2011)}]{2011MUS}%
  \BibitemOpen
  \bibfield  {author} {\bibinfo {author} {\bibfnamefont {M.}~\bibnamefont
  {Musia{\l}}}\ and\ \bibinfo {author} {\bibfnamefont {R.~J.}\ \bibnamefont
  {Bartlett}},\ }\href {\doibase 10.1063/1.3615500} {\bibfield  {journal}
  {\bibinfo  {journal} {J. Chem. Phys.}\ }\textbf {\bibinfo {volume} {135}},\
  \bibinfo {pages} {0} (\bibinfo {year} {2011})}\BibitemShut {NoStop}%
\bibitem [{\citenamefont {Musia}(2012)}]{2012MUS}%
  \BibitemOpen
  \bibfield  {author} {\bibinfo {author} {\bibfnamefont {M.}~\bibnamefont
  {Musia}},\ }\href {\doibase 10.1063/1.3700438} {\bibfield  {journal}
  {\bibinfo  {journal} {J. Chem. Phys.}\ }\textbf {\bibinfo {volume} {136}},\
  \bibinfo {pages} {0} (\bibinfo {year} {2012})}\BibitemShut {NoStop}%
\bibitem [{\citenamefont {Morales}\ \emph {et~al.}(2012)\citenamefont
  {Morales}, \citenamefont {McMinis}, \citenamefont {Clark}, \citenamefont
  {Kim},\ and\ \citenamefont {Scuseria}}]{2012MOR}%
  \BibitemOpen
  \bibfield  {author} {\bibinfo {author} {\bibfnamefont {M.~A.}\ \bibnamefont
  {Morales}}, \bibinfo {author} {\bibfnamefont {J.}~\bibnamefont {McMinis}},
  \bibinfo {author} {\bibfnamefont {B.~K.}\ \bibnamefont {Clark}}, \bibinfo
  {author} {\bibfnamefont {J.}~\bibnamefont {Kim}}, \ and\ \bibinfo {author}
  {\bibfnamefont {G.~E.}\ \bibnamefont {Scuseria}},\ }\href@noop {} {\bibfield
  {journal} {\bibinfo  {journal} {J. Chem. Theory Comput.}\ }\textbf {\bibinfo
  {volume} {8}},\ \bibinfo {pages} {2181} (\bibinfo {year} {2012})}\BibitemShut
  {NoStop}%
\bibitem [{\citenamefont {Reynolds}\ \emph {et~al.}(1982)\citenamefont
  {Reynolds}, \citenamefont {Ceperley}, \citenamefont {Alder},\ and\
  \citenamefont {Lester~Jr}}]{1982REY}%
  \BibitemOpen
  \bibfield  {author} {\bibinfo {author} {\bibfnamefont {P.~J.}\ \bibnamefont
  {Reynolds}}, \bibinfo {author} {\bibfnamefont {D.~M.}\ \bibnamefont
  {Ceperley}}, \bibinfo {author} {\bibfnamefont {B.~J.}\ \bibnamefont {Alder}},
  \ and\ \bibinfo {author} {\bibfnamefont {W.~A.}\ \bibnamefont {Lester~Jr}},\
  }\href@noop {} {\bibfield  {journal} {\bibinfo  {journal} {J. Chem. Phys.}\
  }\textbf {\bibinfo {volume} {77}},\ \bibinfo {pages} {5593} (\bibinfo {year}
  {1982})}\BibitemShut {NoStop}%
\bibitem [{\citenamefont {Reynolds}\ \emph {et~al.}(1986)\citenamefont
  {Reynolds}, \citenamefont {Barnett}, \citenamefont {Hammond},\ and\
  \citenamefont {Lester}}]{1986REY}%
  \BibitemOpen
  \bibfield  {author} {\bibinfo {author} {\bibfnamefont {P.}~\bibnamefont
  {Reynolds}}, \bibinfo {author} {\bibfnamefont {R.}~\bibnamefont {Barnett}},
  \bibinfo {author} {\bibfnamefont {B.}~\bibnamefont {Hammond}}, \ and\
  \bibinfo {author} {\bibfnamefont {W.}~\bibnamefont {Lester}},\ }\href@noop {}
  {\bibfield  {journal} {\bibinfo  {journal} {J. Stat. Phys.}\ }\textbf
  {\bibinfo {volume} {43}},\ \bibinfo {pages} {1017} (\bibinfo {year}
  {1986})}\BibitemShut {NoStop}%
\bibitem [{\citenamefont {Vrbik}\ \emph {et~al.}(1988)\citenamefont {Vrbik},
  \citenamefont {DePasquale},\ and\ \citenamefont {Rothstein}}]{1988VRB}%
  \BibitemOpen
  \bibfield  {author} {\bibinfo {author} {\bibfnamefont {J.}~\bibnamefont
  {Vrbik}}, \bibinfo {author} {\bibfnamefont {M.~F.}\ \bibnamefont
  {DePasquale}}, \ and\ \bibinfo {author} {\bibfnamefont {S.~M.}\ \bibnamefont
  {Rothstein}},\ }\href@noop {} {\bibfield  {journal} {\bibinfo  {journal} {J.
  Chem. Phys.}\ }\textbf {\bibinfo {volume} {88}},\ \bibinfo {pages} {3784}
  (\bibinfo {year} {1988})}\BibitemShut {NoStop}%
\bibitem [{\citenamefont {Lester~Jr}\ and\ \citenamefont
  {Hammond}(1990)}]{1990LES}%
  \BibitemOpen
  \bibfield  {author} {\bibinfo {author} {\bibfnamefont {W.~A.}\ \bibnamefont
  {Lester~Jr}}\ and\ \bibinfo {author} {\bibfnamefont {B.~L.}\ \bibnamefont
  {Hammond}},\ }\href@noop {} {\bibfield  {journal} {\bibinfo  {journal} {Annu.
  Rev. Phys. Chem.}\ }\textbf {\bibinfo {volume} {41}},\ \bibinfo {pages} {283}
  (\bibinfo {year} {1990})}\BibitemShut {NoStop}%
\bibitem [{\citenamefont {Kenny}\ \emph {et~al.}(1995)\citenamefont {Kenny},
  \citenamefont {Rajagopal},\ and\ \citenamefont {Needs}}]{1995KEN}%
  \BibitemOpen
  \bibfield  {author} {\bibinfo {author} {\bibfnamefont {S.}~\bibnamefont
  {Kenny}}, \bibinfo {author} {\bibfnamefont {G.}~\bibnamefont {Rajagopal}}, \
  and\ \bibinfo {author} {\bibfnamefont {R.}~\bibnamefont {Needs}},\
  }\href@noop {} {\bibfield  {journal} {\bibinfo  {journal} {Phys. Rev. A}\
  }\textbf {\bibinfo {volume} {51}},\ \bibinfo {pages} {1898} (\bibinfo {year}
  {1995})}\BibitemShut {NoStop}%
\bibitem [{\citenamefont {L{\"u}chow}\ and\ \citenamefont
  {Anderson}(1996)}]{1996LUC}%
  \BibitemOpen
  \bibfield  {author} {\bibinfo {author} {\bibfnamefont {A.}~\bibnamefont
  {L{\"u}chow}}\ and\ \bibinfo {author} {\bibfnamefont {J.~B.}\ \bibnamefont
  {Anderson}},\ }\href@noop {} {\bibfield  {journal} {\bibinfo  {journal} {J.
  Chem. Phys.}\ }\textbf {\bibinfo {volume} {105}},\ \bibinfo {pages} {7573}
  (\bibinfo {year} {1996})}\BibitemShut {NoStop}%
\bibitem [{\citenamefont {Yoshida}\ and\ \citenamefont
  {Miyako}(1997)}]{1997YOS}%
  \BibitemOpen
  \bibfield  {author} {\bibinfo {author} {\bibfnamefont {T.}~\bibnamefont
  {Yoshida}}\ and\ \bibinfo {author} {\bibfnamefont {G.}~\bibnamefont
  {Miyako}},\ }\href@noop {} {\bibfield  {journal} {\bibinfo  {journal} {J.
  Chem. Phys.}\ }\textbf {\bibinfo {volume} {107}},\ \bibinfo {pages} {3864}
  (\bibinfo {year} {1997})}\BibitemShut {NoStop}%
\bibitem [{\citenamefont {Huang}\ \emph {et~al.}(1997)\citenamefont {Huang},
  \citenamefont {Umrigar},\ and\ \citenamefont {Nightingale}}]{1997HUA}%
  \BibitemOpen
  \bibfield  {author} {\bibinfo {author} {\bibfnamefont {C.-J.}\ \bibnamefont
  {Huang}}, \bibinfo {author} {\bibfnamefont {C.}~\bibnamefont {Umrigar}}, \
  and\ \bibinfo {author} {\bibfnamefont {M.}~\bibnamefont {Nightingale}},\
  }\href@noop {} {\bibfield  {journal} {\bibinfo  {journal} {J. Chem. Phys.}\
  }\textbf {\bibinfo {volume} {107}},\ \bibinfo {pages} {3007} (\bibinfo {year}
  {1997})}\BibitemShut {NoStop}%
\bibitem [{\citenamefont {Shlyakhter}\ \emph {et~al.}(1999)\citenamefont
  {Shlyakhter}, \citenamefont {Sokolova}, \citenamefont {L{\"u}chow},\ and\
  \citenamefont {Anderson}}]{1999SHL}%
  \BibitemOpen
  \bibfield  {author} {\bibinfo {author} {\bibfnamefont {Y.}~\bibnamefont
  {Shlyakhter}}, \bibinfo {author} {\bibfnamefont {S.}~\bibnamefont
  {Sokolova}}, \bibinfo {author} {\bibfnamefont {A.}~\bibnamefont
  {L{\"u}chow}}, \ and\ \bibinfo {author} {\bibfnamefont {J.~B.}\ \bibnamefont
  {Anderson}},\ }\href@noop {} {\bibfield  {journal} {\bibinfo  {journal} {J.
  Chem. Phys.}\ }\textbf {\bibinfo {volume} {110}},\ \bibinfo {pages} {10725}
  (\bibinfo {year} {1999})}\BibitemShut {NoStop}%
\bibitem [{\citenamefont {Sarsa}\ \emph {et~al.}(2002)\citenamefont {Sarsa},
  \citenamefont {Boronat},\ and\ \citenamefont {Casulleras}}]{2002SAR}%
  \BibitemOpen
  \bibfield  {author} {\bibinfo {author} {\bibfnamefont {A.}~\bibnamefont
  {Sarsa}}, \bibinfo {author} {\bibfnamefont {J.}~\bibnamefont {Boronat}}, \
  and\ \bibinfo {author} {\bibfnamefont {J.}~\bibnamefont {Casulleras}},\
  }\href@noop {} {\bibfield  {journal} {\bibinfo  {journal} {J. Chem. Phys.}\
  }\textbf {\bibinfo {volume} {116}},\ \bibinfo {pages} {5956} (\bibinfo {year}
  {2002})}\BibitemShut {NoStop}%
\bibitem [{\citenamefont {Casula}\ and\ \citenamefont
  {Sorella}(2003)}]{2003CAS}%
  \BibitemOpen
  \bibfield  {author} {\bibinfo {author} {\bibfnamefont {M.}~\bibnamefont
  {Casula}}\ and\ \bibinfo {author} {\bibfnamefont {S.}~\bibnamefont
  {Sorella}},\ }\href@noop {} {\bibfield  {journal} {\bibinfo  {journal} {J.
  Chem. Phys.}\ }\textbf {\bibinfo {volume} {119}},\ \bibinfo {pages} {6500}
  (\bibinfo {year} {2003})}\BibitemShut {NoStop}%
\bibitem [{\citenamefont {Casula}\ \emph {et~al.}(2004)\citenamefont {Casula},
  \citenamefont {Attaccalite},\ and\ \citenamefont {Sorella}}]{2004CAS}%
  \BibitemOpen
  \bibfield  {author} {\bibinfo {author} {\bibfnamefont {M.}~\bibnamefont
  {Casula}}, \bibinfo {author} {\bibfnamefont {C.}~\bibnamefont {Attaccalite}},
  \ and\ \bibinfo {author} {\bibfnamefont {S.}~\bibnamefont {Sorella}},\ }\href
  {\doibase 10.1063/1.1794632} {\bibfield  {journal} {\bibinfo  {journal} {J.
  Chem. Phys.}\ }\textbf {\bibinfo {volume} {121}},\ \bibinfo {pages} {7110}
  (\bibinfo {year} {2004})}\BibitemShut {NoStop}%
\bibitem [{\citenamefont {Caffarel}\ \emph {et~al.}(2005)\citenamefont
  {Caffarel}, \citenamefont {Daudey}, \citenamefont {Heully},\ and\
  \citenamefont {Ram{\'\i}rez-Sol{\'\i}s}}]{2005CAF}%
  \BibitemOpen
  \bibfield  {author} {\bibinfo {author} {\bibfnamefont {M.}~\bibnamefont
  {Caffarel}}, \bibinfo {author} {\bibfnamefont {J.-P.}\ \bibnamefont
  {Daudey}}, \bibinfo {author} {\bibfnamefont {J.-L.}\ \bibnamefont {Heully}},
  \ and\ \bibinfo {author} {\bibfnamefont {A.}~\bibnamefont
  {Ram{\'\i}rez-Sol{\'\i}s}},\ }\href@noop {} {\bibfield  {journal} {\bibinfo
  {journal} {J. Chem. Phys.}\ }\textbf {\bibinfo {volume} {123}},\ \bibinfo
  {pages} {094102} (\bibinfo {year} {2005})}\BibitemShut {NoStop}%
\bibitem [{\citenamefont {Buend{\'\i}a}\ \emph {et~al.}(2006)\citenamefont
  {Buend{\'\i}a}, \citenamefont {G{\'a}lvez},\ and\ \citenamefont
  {Sarsa}}]{2006BUE}%
  \BibitemOpen
  \bibfield  {author} {\bibinfo {author} {\bibfnamefont {E.}~\bibnamefont
  {Buend{\'\i}a}}, \bibinfo {author} {\bibfnamefont {F.}~\bibnamefont
  {G{\'a}lvez}}, \ and\ \bibinfo {author} {\bibfnamefont {A.}~\bibnamefont
  {Sarsa}},\ }\href@noop {} {\bibfield  {journal} {\bibinfo  {journal} {Chem.
  Phys. Lett.}\ }\textbf {\bibinfo {volume} {428}},\ \bibinfo {pages} {241}
  (\bibinfo {year} {2006})}\BibitemShut {NoStop}%
\bibitem [{\citenamefont {Nemec}\ \emph {et~al.}(2010)\citenamefont {Nemec},
  \citenamefont {Towler},\ and\ \citenamefont {Needs}}]{2010NEM}%
  \BibitemOpen
  \bibfield  {author} {\bibinfo {author} {\bibfnamefont {N.}~\bibnamefont
  {Nemec}}, \bibinfo {author} {\bibfnamefont {M.~D.}\ \bibnamefont {Towler}}, \
  and\ \bibinfo {author} {\bibfnamefont {R.}~\bibnamefont {Needs}},\
  }\href@noop {} {\bibfield  {journal} {\bibinfo  {journal} {J. Chem. Phys.}\
  }\textbf {\bibinfo {volume} {132}},\ \bibinfo {pages} {034111} (\bibinfo
  {year} {2010})}\BibitemShut {NoStop}%
\bibitem [{\citenamefont {Scemama}\ \emph {et~al.}(2014)\citenamefont
  {Scemama}, \citenamefont {Applencourt}, \citenamefont {Giner},\ and\
  \citenamefont {Caffarel}}]{2014SCE}%
  \BibitemOpen
  \bibfield  {author} {\bibinfo {author} {\bibfnamefont {A.}~\bibnamefont
  {Scemama}}, \bibinfo {author} {\bibfnamefont {T.}~\bibnamefont
  {Applencourt}}, \bibinfo {author} {\bibfnamefont {E.}~\bibnamefont {Giner}},
  \ and\ \bibinfo {author} {\bibfnamefont {M.}~\bibnamefont {Caffarel}},\
  }\href@noop {} {\bibfield  {journal} {\bibinfo  {journal} {J. Chem. Phys.}\
  }\textbf {\bibinfo {volume} {141}},\ \bibinfo {pages} {244110} (\bibinfo
  {year} {2014})}\BibitemShut {NoStop}%
\bibitem [{\citenamefont {Powell}\ and\ \citenamefont {Dawes}(2016)}]{2016POW}%
  \BibitemOpen
  \bibfield  {author} {\bibinfo {author} {\bibfnamefont {A.~D.}\ \bibnamefont
  {Powell}}\ and\ \bibinfo {author} {\bibfnamefont {R.}~\bibnamefont {Dawes}},\
  }\href@noop {} {\bibfield  {journal} {\bibinfo  {journal} {J. Chem. Phys.}\
  }\textbf {\bibinfo {volume} {145}},\ \bibinfo {pages} {224308} (\bibinfo
  {year} {2016})}\BibitemShut {NoStop}%
\bibitem [{\citenamefont {Maezono}\ \emph {et~al.}(2003)\citenamefont
  {Maezono}, \citenamefont {Towler}, \citenamefont {Lee},\ and\ \citenamefont
  {Needs}}]{2003MAE}%
  \BibitemOpen
  \bibfield  {author} {\bibinfo {author} {\bibfnamefont {R.}~\bibnamefont
  {Maezono}}, \bibinfo {author} {\bibfnamefont {M.}~\bibnamefont {Towler}},
  \bibinfo {author} {\bibfnamefont {Y.}~\bibnamefont {Lee}}, \ and\ \bibinfo
  {author} {\bibfnamefont {R.}~\bibnamefont {Needs}},\ }\href@noop {}
  {\bibfield  {journal} {\bibinfo  {journal} {Phys. Rev. B}\ }\textbf {\bibinfo
  {volume} {68}},\ \bibinfo {pages} {165103} (\bibinfo {year}
  {2003})}\BibitemShut {NoStop}%
\bibitem [{\citenamefont {Feller}(1996)}]{1996FEL}%
  \BibitemOpen
  \bibfield  {author} {\bibinfo {author} {\bibfnamefont {D.}~\bibnamefont
  {Feller}},\ }\href {\doibase
  10.1002/(SICI)1096-987X(199610)17:13<1571::AID-JCC9>3.0.CO;2-P} {\bibfield
  {journal} {\bibinfo  {journal} {J. Comput. Chem.}\ }\textbf {\bibinfo
  {volume} {17}},\ \bibinfo {pages} {1571} (\bibinfo {year}
  {1996})}\BibitemShut {NoStop}%
\bibitem [{\citenamefont {Schuchardt}\ \emph {et~al.}(2007)\citenamefont
  {Schuchardt}, \citenamefont {Didier}, \citenamefont {Elsethagen},
  \citenamefont {Sun}, \citenamefont {Gurumoorthi}, \citenamefont {Chase},
  \citenamefont {Li},\ and\ \citenamefont {Windus}}]{2007SCU}%
  \BibitemOpen
  \bibfield  {author} {\bibinfo {author} {\bibfnamefont {K.~L.}\ \bibnamefont
  {Schuchardt}}, \bibinfo {author} {\bibfnamefont {B.~T.}\ \bibnamefont
  {Didier}}, \bibinfo {author} {\bibfnamefont {T.}~\bibnamefont {Elsethagen}},
  \bibinfo {author} {\bibfnamefont {L.}~\bibnamefont {Sun}}, \bibinfo {author}
  {\bibfnamefont {V.}~\bibnamefont {Gurumoorthi}}, \bibinfo {author}
  {\bibfnamefont {J.}~\bibnamefont {Chase}}, \bibinfo {author} {\bibfnamefont
  {J.}~\bibnamefont {Li}}, \ and\ \bibinfo {author} {\bibfnamefont {T.~L.}\
  \bibnamefont {Windus}},\ }\href {\doibase 10.1021/CI600510J} {\bibfield
  {journal} {\bibinfo  {journal} {J. Chem. Inf. Model.}\ }\textbf {\bibinfo
  {volume} {47}},\ \bibinfo {pages} {1045} (\bibinfo {year}
  {2007})}\BibitemShut {NoStop}%
\bibitem [{\citenamefont {Sorella}\ \emph {et~al.}(2007)\citenamefont
  {Sorella}, \citenamefont {Casula},\ and\ \citenamefont {Rocca}}]{benzene}%
  \BibitemOpen
  \bibfield  {author} {\bibinfo {author} {\bibfnamefont {S.}~\bibnamefont
  {Sorella}}, \bibinfo {author} {\bibfnamefont {M.}~\bibnamefont {Casula}}, \
  and\ \bibinfo {author} {\bibfnamefont {D.}~\bibnamefont {Rocca}},\
  }\href@noop {} {\bibfield  {journal} {\bibinfo  {journal} {J. Chem. Phys.}\
  }\textbf {\bibinfo {volume} {127}},\ \bibinfo {pages} {014105} (\bibinfo
  {year} {2007})}\BibitemShut {NoStop}%
\bibitem [{\citenamefont {Umrigar}\ \emph {et~al.}(2007)\citenamefont
  {Umrigar}, \citenamefont {Toulouse}, \citenamefont {Filippi}, \citenamefont
  {Sorella},\ and\ \citenamefont {Rhenning}}]{linear}%
  \BibitemOpen
  \bibfield  {author} {\bibinfo {author} {\bibfnamefont {C.~J.}\ \bibnamefont
  {Umrigar}}, \bibinfo {author} {\bibfnamefont {J.}~\bibnamefont {Toulouse}},
  \bibinfo {author} {\bibfnamefont {C.}~\bibnamefont {Filippi}}, \bibinfo
  {author} {\bibfnamefont {S.}~\bibnamefont {Sorella}}, \ and\ \bibinfo
  {author} {\bibfnamefont {H.}~\bibnamefont {Rhenning}},\ }\href@noop {}
  {\bibfield  {journal} {\bibinfo  {journal} {Phys. Rev. Lett}\ }\textbf
  {\bibinfo {volume} {98}},\ \bibinfo {pages} {110201} (\bibinfo {year}
  {2007})}\BibitemShut {NoStop}%
\bibitem [{\citenamefont {{Ten Haaf}}\ \emph {et~al.}(1995)\citenamefont {{Ten
  Haaf}}, \citenamefont {{Van Bemmel}}, \citenamefont {{Van Leeuwen}},
  \citenamefont {{Van Saarloos}},\ and\ \citenamefont {Ceperley}}]{1994TEN}%
  \BibitemOpen
  \bibfield  {author} {\bibinfo {author} {\bibfnamefont {D.~F.}\ \bibnamefont
  {{Ten Haaf}}}, \bibinfo {author} {\bibfnamefont {H.~J.}\ \bibnamefont {{Van
  Bemmel}}}, \bibinfo {author} {\bibfnamefont {J.~M.}\ \bibnamefont {{Van
  Leeuwen}}}, \bibinfo {author} {\bibfnamefont {W.}~\bibnamefont {{Van
  Saarloos}}}, \ and\ \bibinfo {author} {\bibfnamefont {D.~M.}\ \bibnamefont
  {Ceperley}},\ }\href {\doibase 10.1103/PhysRevB.51.13039} {\bibfield
  {journal} {\bibinfo  {journal} {Phys. Rev. B}\ }\textbf {\bibinfo {volume}
  {51}},\ \bibinfo {pages} {13039} (\bibinfo {year} {1995})}\BibitemShut
  {NoStop}%
\bibitem [{\citenamefont {{Calandra Buonaura}}\ and\ \citenamefont
  {Sorella}(1998)}]{1998BUO}%
  \BibitemOpen
  \bibfield  {author} {\bibinfo {author} {\bibfnamefont {M.}~\bibnamefont
  {{Calandra Buonaura}}}\ and\ \bibinfo {author} {\bibfnamefont
  {S.}~\bibnamefont {Sorella}},\ }\href {\doibase 10.1103/PhysRevB.57.11446}
  {\bibfield  {journal} {\bibinfo  {journal} {Phys. Rev. B}\ }\textbf {\bibinfo
  {volume} {57}},\ \bibinfo {pages} {11446} (\bibinfo {year}
  {1998})}\BibitemShut {NoStop}%
\bibitem [{\citenamefont {Sorella}\ and\ \citenamefont
  {Capriotti}(2000)}]{2000SOR}%
  \BibitemOpen
  \bibfield  {author} {\bibinfo {author} {\bibfnamefont {S.}~\bibnamefont
  {Sorella}}\ and\ \bibinfo {author} {\bibfnamefont {L.}~\bibnamefont
  {Capriotti}},\ }\href {\doibase 10.1103/PhysRevB.61.2599} {\bibfield
  {journal} {\bibinfo  {journal} {Phys. Rev. B}\ }\textbf {\bibinfo {volume}
  {61}},\ \bibinfo {pages} {2599} (\bibinfo {year} {2000})}\BibitemShut
  {NoStop}%
\bibitem [{\citenamefont {Casula}\ \emph {et~al.}(2005)\citenamefont {Casula},
  \citenamefont {Filippi},\ and\ \citenamefont {Sorella}}]{2005CAS}%
  \BibitemOpen
  \bibfield  {author} {\bibinfo {author} {\bibfnamefont {M.}~\bibnamefont
  {Casula}}, \bibinfo {author} {\bibfnamefont {C.}~\bibnamefont {Filippi}}, \
  and\ \bibinfo {author} {\bibfnamefont {S.}~\bibnamefont {Sorella}},\
  }\href@noop {} {\bibfield  {journal} {\bibinfo  {journal} {Phys. Rev. Lett.}\
  }\textbf {\bibinfo {volume} {95}},\ \bibinfo {pages} {1} (\bibinfo {year}
  {2005})}\BibitemShut {NoStop}%
\bibitem [{\citenamefont {Casula}\ \emph {et~al.}(2006)\citenamefont {Casula},
  \citenamefont {Sorella},\ and\ \citenamefont {Senatore}}]{2006CAS}%
  \BibitemOpen
  \bibfield  {author} {\bibinfo {author} {\bibfnamefont {M.}~\bibnamefont
  {Casula}}, \bibinfo {author} {\bibfnamefont {S.}~\bibnamefont {Sorella}}, \
  and\ \bibinfo {author} {\bibfnamefont {G.}~\bibnamefont {Senatore}},\ }\href
  {\doibase 10.1103/PhysRevB.74.245427} {\bibfield  {journal} {\bibinfo
  {journal} {Phys. Rev. B}\ }\textbf {\bibinfo {volume} {74}},\ \bibinfo
  {pages} {1} (\bibinfo {year} {2006})},\ \Eprint
  {http://arxiv.org/abs/0607130} {0607130} \BibitemShut {NoStop}%
\bibitem [{\citenamefont {Casula}\ \emph {et~al.}(2010)\citenamefont {Casula},
  \citenamefont {Moroni}, \citenamefont {Sorella},\ and\ \citenamefont
  {Filippi}}]{2010CAS}%
  \BibitemOpen
  \bibfield  {author} {\bibinfo {author} {\bibfnamefont {M.}~\bibnamefont
  {Casula}}, \bibinfo {author} {\bibfnamefont {S.}~\bibnamefont {Moroni}},
  \bibinfo {author} {\bibfnamefont {S.}~\bibnamefont {Sorella}}, \ and\
  \bibinfo {author} {\bibfnamefont {C.}~\bibnamefont {Filippi}},\ }\href@noop
  {} {\bibfield  {journal} {\bibinfo  {journal} {J. Chem. Phys.}\ }\textbf
  {\bibinfo {volume} {132}},\ \bibinfo {pages} {1} (\bibinfo {year}
  {2010})}\BibitemShut {NoStop}%
\bibitem [{\citenamefont {Becca}\ and\ \citenamefont
  {Sorella}(2017)}]{2017BEC}%
  \BibitemOpen
  \bibfield  {author} {\bibinfo {author} {\bibfnamefont {F.}~\bibnamefont
  {Becca}}\ and\ \bibinfo {author} {\bibfnamefont {S.}~\bibnamefont
  {Sorella}},\ }\href@noop {} {\emph {\bibinfo {title} {{Quantum Monte Carlo
  approaches for correlated systems}}}}\ (\bibinfo  {publisher} {Cambridge
  University Press},\ \bibinfo {year} {2017})\BibitemShut {NoStop}%
\bibitem [{\citenamefont {Ono}\ and\ \citenamefont {Hirose}(1999)}]{1999ONO}%
  \BibitemOpen
  \bibfield  {author} {\bibinfo {author} {\bibfnamefont {T.}~\bibnamefont
  {Ono}}\ and\ \bibinfo {author} {\bibfnamefont {K.}~\bibnamefont {Hirose}},\
  }\href@noop {} {\bibfield  {journal} {\bibinfo  {journal} {Phys. Rev. Lett.}\
  }\textbf {\bibinfo {volume} {82}},\ \bibinfo {pages} {5016} (\bibinfo {year}
  {1999})}\BibitemShut {NoStop}%
\bibitem [{\citenamefont {Ono}\ and\ \citenamefont {Hirose}(2005)}]{2005ONO}%
  \BibitemOpen
  \bibfield  {author} {\bibinfo {author} {\bibfnamefont {T.}~\bibnamefont
  {Ono}}\ and\ \bibinfo {author} {\bibfnamefont {K.}~\bibnamefont {Hirose}},\
  }\href {https://journals.aps.org/prb/pdf/10.1103/PhysRevB.72.085115}
  {\bibfield  {journal} {\bibinfo  {journal} {Phys. Rev. B}\ }\textbf {\bibinfo
  {volume} {72}} (\bibinfo {year} {2005})}\BibitemShut {NoStop}%
\bibitem [{\citenamefont {Perdew}\ and\ \citenamefont
  {Zunger}(1981)}]{1981PER}%
  \BibitemOpen
  \bibfield  {author} {\bibinfo {author} {\bibfnamefont {J.~P.}\ \bibnamefont
  {Perdew}}\ and\ \bibinfo {author} {\bibfnamefont {A.}~\bibnamefont
  {Zunger}},\ }\href {\doibase 10.1103/PhysRevB.23.5048} {\bibfield  {journal}
  {\bibinfo  {journal} {Phys. Rev. B}\ }\textbf {\bibinfo {volume} {23}},\
  \bibinfo {pages} {5048} (\bibinfo {year} {1981})}\BibitemShut {NoStop}%
\bibitem [{\citenamefont {Chakravorty}\ \emph {et~al.}(1993)\citenamefont
  {Chakravorty}, \citenamefont {Gwaltney}, \citenamefont {Davidson},
  \citenamefont {Parpia},\ and\ \citenamefont {p~Fischer}}]{1993CHA}%
  \BibitemOpen
  \bibfield  {author} {\bibinfo {author} {\bibfnamefont {S.~J.}\ \bibnamefont
  {Chakravorty}}, \bibinfo {author} {\bibfnamefont {S.~R.}\ \bibnamefont
  {Gwaltney}}, \bibinfo {author} {\bibfnamefont {E.~R.}\ \bibnamefont
  {Davidson}}, \bibinfo {author} {\bibfnamefont {F.~A.}\ \bibnamefont
  {Parpia}}, \ and\ \bibinfo {author} {\bibfnamefont {C.~F.}\ \bibnamefont
  {p~Fischer}},\ }\href {\doibase 10.1103/PhysRevA.47.3649} {\bibfield
  {journal} {\bibinfo  {journal} {Phys. Rev. A}\ }\textbf {\bibinfo {volume}
  {47}},\ \bibinfo {pages} {3649} (\bibinfo {year} {1993})}\BibitemShut
  {NoStop}%
\bibitem [{Note1()}]{Note1}%
  \BibitemOpen
  \bibinfo {note} {The aug-cc-VQZ basis sets were used for these
  calculations.}\BibitemShut {Stop}%
\bibitem [{\citenamefont {Frisch}\ \emph {et~al.}(2013)\citenamefont {Frisch},
  \citenamefont {Trucks}, \citenamefont {Schlegel}, \citenamefont {Scuseria},
  \citenamefont {Robb}, \citenamefont {Cheeseman}, \citenamefont {Scalmani},
  \citenamefont {Mennucci}, \citenamefont {Petersson} \emph
  {et~al.}}]{2013FRI}%
  \BibitemOpen
  \bibfield  {author} {\bibinfo {author} {\bibfnamefont {M.}~\bibnamefont
  {Frisch}}, \bibinfo {author} {\bibfnamefont {G.}~\bibnamefont {Trucks}},
  \bibinfo {author} {\bibfnamefont {H.}~\bibnamefont {Schlegel}}, \bibinfo
  {author} {\bibfnamefont {G.}~\bibnamefont {Scuseria}}, \bibinfo {author}
  {\bibfnamefont {M.}~\bibnamefont {Robb}}, \bibinfo {author} {\bibfnamefont
  {J.}~\bibnamefont {Cheeseman}}, \bibinfo {author} {\bibfnamefont
  {V.}~\bibnamefont {Scalmani}, \bibfnamefont {G~.and~Barone}}, \bibinfo
  {author} {\bibfnamefont {B.}~\bibnamefont {Mennucci}}, \bibinfo {author}
  {\bibfnamefont {G.}~\bibnamefont {Petersson}},  \emph {et~al.},\ }\href@noop
  {} {\emph {\bibinfo {title} {{Gaussian 09, Revision E.01}}}}\ (\bibinfo
  {publisher} {Gaussian Inc.},\ \bibinfo {address} {Wallingford CT},\ \bibinfo
  {year} {2013})\BibitemShut {NoStop}%
\bibitem [{\citenamefont {Sorbie}\ and\ \citenamefont
  {Murrell}(1975)}]{1975SOR}%
  \BibitemOpen
  \bibfield  {author} {\bibinfo {author} {\bibfnamefont {K.}~\bibnamefont
  {Sorbie}}\ and\ \bibinfo {author} {\bibfnamefont {J.}~\bibnamefont
  {Murrell}},\ }\href@noop {} {\bibfield  {journal} {\bibinfo  {journal} {Mol.
  Phys.}\ }\textbf {\bibinfo {volume} {29}},\ \bibinfo {pages} {1387} (\bibinfo
  {year} {1975})}\BibitemShut {NoStop}%
\bibitem [{\citenamefont {Jones}\ \emph {et~al.}()\citenamefont {Jones},
  \citenamefont {Oliphant}, \citenamefont {Peterson} \emph {et~al.}}]{2001SCI}%
  \BibitemOpen
  \bibfield  {author} {\bibinfo {author} {\bibfnamefont {E.}~\bibnamefont
  {Jones}}, \bibinfo {author} {\bibfnamefont {T.}~\bibnamefont {Oliphant}},
  \bibinfo {author} {\bibfnamefont {P.}~\bibnamefont {Peterson}},  \emph
  {et~al.},\ }\href@noop {} {\enquote {\bibinfo {title} {{SciPy}: Open source
  scientific tools for {Python}},}\ }\bibinfo {howpublished}
  {\url{http://www.scipy.org}},\ \bibinfo {note} {[Online; accessed
  10-March-2019]}\BibitemShut {NoStop}%
\bibitem [{\citenamefont {Marchi}\ \emph {et~al.}(2009)\citenamefont {Marchi},
  \citenamefont {Azadi}, \citenamefont {Casula},\ and\ \citenamefont
  {Sorella}}]{2009MAR}%
  \BibitemOpen
  \bibfield  {author} {\bibinfo {author} {\bibfnamefont {M.}~\bibnamefont
  {Marchi}}, \bibinfo {author} {\bibfnamefont {S.}~\bibnamefont {Azadi}},
  \bibinfo {author} {\bibfnamefont {M.}~\bibnamefont {Casula}}, \ and\ \bibinfo
  {author} {\bibfnamefont {S.}~\bibnamefont {Sorella}},\ }\href@noop {}
  {\bibfield  {journal} {\bibinfo  {journal} {J. Chem. Phys.}\ }\textbf
  {\bibinfo {volume} {131}},\ \bibinfo {pages} {154116} (\bibinfo {year}
  {2009})}\BibitemShut {NoStop}%
\bibitem [{\citenamefont {Liu}\ \emph {et~al.}(1995)\citenamefont {Liu},
  \citenamefont {Carter},\ and\ \citenamefont {Carter}}]{1995LIU}%
  \BibitemOpen
  \bibfield  {author} {\bibinfo {author} {\bibfnamefont {Z.}~\bibnamefont
  {Liu}}, \bibinfo {author} {\bibfnamefont {L.~E.}\ \bibnamefont {Carter}}, \
  and\ \bibinfo {author} {\bibfnamefont {E.~A.}\ \bibnamefont {Carter}},\
  }\href@noop {} {\bibfield  {journal} {\bibinfo  {journal} {J. Phys. Chem.}\
  }\textbf {\bibinfo {volume} {99}},\ \bibinfo {pages} {4355} (\bibinfo {year}
  {1995})}\BibitemShut {NoStop}%
\bibitem [{\citenamefont {Huber}(2013)}]{2013HUB}%
  \BibitemOpen
  \bibfield  {author} {\bibinfo {author} {\bibfnamefont {K.-P.}\ \bibnamefont
  {Huber}},\ }\href {\doibase 10.1119/1.1932852} {\emph {\bibinfo {title}
  {{Molecular spectra and molecular structure: IV. Constants of diatomic
  molecules}}}}\ (\bibinfo  {publisher} {Springer Science \& Business Media},\
  \bibinfo {year} {2013})\BibitemShut {NoStop}%
\bibitem [{Note2()}]{Note2}%
  \BibitemOpen
  \bibinfo {note} {The original value is 14.918$\pm $0.357 kcal/mol at $d$ =
  3.0789 \r A. See their supplementary material.}\BibitemShut {Stop}%
\bibitem [{\citenamefont {Zen}\ \emph {et~al.}(2014)\citenamefont {Zen},
  \citenamefont {Coccia}, \citenamefont {Luo}, \citenamefont {Sorella},\ and\
  \citenamefont {Guidoni}}]{2014ZEN}%
  \BibitemOpen
  \bibfield  {author} {\bibinfo {author} {\bibfnamefont {A.}~\bibnamefont
  {Zen}}, \bibinfo {author} {\bibfnamefont {E.}~\bibnamefont {Coccia}},
  \bibinfo {author} {\bibfnamefont {Y.}~\bibnamefont {Luo}}, \bibinfo {author}
  {\bibfnamefont {S.}~\bibnamefont {Sorella}}, \ and\ \bibinfo {author}
  {\bibfnamefont {L.}~\bibnamefont {Guidoni}},\ }\href {\doibase
  10.1021/ct401008s} {\bibfield  {journal} {\bibinfo  {journal} {J. Chem.
  Theory Comput.}\ }\textbf {\bibinfo {volume} {10}},\ \bibinfo {pages} {1048}
  (\bibinfo {year} {2014})}\BibitemShut {NoStop}%
\bibitem [{\citenamefont {Genovese}\ \emph {et~al.}(2019)\citenamefont
  {Genovese}, \citenamefont {Meninno},\ and\ \citenamefont
  {Sorella}}]{2019GEN}%
  \BibitemOpen
  \bibfield  {author} {\bibinfo {author} {\bibfnamefont {C.}~\bibnamefont
  {Genovese}}, \bibinfo {author} {\bibfnamefont {A.}~\bibnamefont {Meninno}}, \
  and\ \bibinfo {author} {\bibfnamefont {S.}~\bibnamefont {Sorella}},\ }\href
  {\doibase 10.1063/1.5081933} {\bibfield  {journal} {\bibinfo  {journal} {J.
  Chem. Phys.}\ }\textbf {\bibinfo {volume} {150}},\ \bibinfo {pages} {084102}
  (\bibinfo {year} {2019})}\BibitemShut {NoStop}%
\bibitem [{\citenamefont {{L{\'{o}}pez R{\'{i}}os}}\ \emph
  {et~al.}(2006)\citenamefont {{L{\'{o}}pez R{\'{i}}os}}, \citenamefont {Ma},
  \citenamefont {Drummond}, \citenamefont {Towler},\ and\ \citenamefont
  {Needs}}]{2006ROP}%
  \BibitemOpen
  \bibfield  {author} {\bibinfo {author} {\bibfnamefont {P.}~\bibnamefont
  {{L{\'{o}}pez R{\'{i}}os}}}, \bibinfo {author} {\bibfnamefont
  {A.}~\bibnamefont {Ma}}, \bibinfo {author} {\bibfnamefont {N.~D.}\
  \bibnamefont {Drummond}}, \bibinfo {author} {\bibfnamefont {M.~D.}\
  \bibnamefont {Towler}}, \ and\ \bibinfo {author} {\bibfnamefont {R.~J.}\
  \bibnamefont {Needs}},\ }\href
  {https://journals.aps.org/pre/pdf/10.1103/PhysRevE.74.066701} {\bibfield
  {journal} {\bibinfo  {journal} {Phys. Rev. E}\ }\textbf {\bibinfo {volume}
  {74}} (\bibinfo {year} {2006})}\BibitemShut {NoStop}%
\bibitem [{\citenamefont {Bajdich}\ \emph {et~al.}(2006)\citenamefont
  {Bajdich}, \citenamefont {Mitas}, \citenamefont {Drobn{\'{y}}}, \citenamefont
  {Wagner},\ and\ \citenamefont {Schmidt}}]{2006BAJ}%
  \BibitemOpen
  \bibfield  {author} {\bibinfo {author} {\bibfnamefont {M.}~\bibnamefont
  {Bajdich}}, \bibinfo {author} {\bibfnamefont {L.}~\bibnamefont {Mitas}},
  \bibinfo {author} {\bibfnamefont {G.}~\bibnamefont {Drobn{\'{y}}}}, \bibinfo
  {author} {\bibfnamefont {L.~K.}\ \bibnamefont {Wagner}}, \ and\ \bibinfo
  {author} {\bibfnamefont {K.~E.}\ \bibnamefont {Schmidt}},\ }\href
  {https://journals.aps.org/prl/pdf/10.1103/PhysRevLett.96.130201} {\bibfield
  {journal} {\bibinfo  {journal} {Phys. Rev. Lett.}\ }\textbf {\bibinfo
  {volume} {96}} (\bibinfo {year} {2006})}\BibitemShut {NoStop}%
\bibitem [{\citenamefont {Momma}\ and\ \citenamefont {Izumi}(2011)}]{2011MOM}%
  \BibitemOpen
  \bibfield  {author} {\bibinfo {author} {\bibfnamefont {K.}~\bibnamefont
  {Momma}}\ and\ \bibinfo {author} {\bibfnamefont {F.}~\bibnamefont {Izumi}},\
  }\href {\doibase 10.1107/S0021889811038970} {\bibfield  {journal} {\bibinfo
  {journal} {J. Appl. Crystallogr.}\ }\textbf {\bibinfo {volume} {44}},\
  \bibinfo {pages} {1272} (\bibinfo {year} {2011})}\BibitemShut {NoStop}%
\bibitem [{\citenamefont {Sch{\"a}fer}\ \emph {et~al.}(1992)\citenamefont
  {Sch{\"a}fer}, \citenamefont {Horn},\ and\ \citenamefont
  {Ahlrichs}}]{1992SCH}%
  \BibitemOpen
  \bibfield  {author} {\bibinfo {author} {\bibfnamefont {A.}~\bibnamefont
  {Sch{\"a}fer}}, \bibinfo {author} {\bibfnamefont {H.}~\bibnamefont {Horn}}, \
  and\ \bibinfo {author} {\bibfnamefont {R.}~\bibnamefont {Ahlrichs}},\
  }\href@noop {} {\bibfield  {journal} {\bibinfo  {journal} {J. Chem. Phys.}\
  }\textbf {\bibinfo {volume} {97}},\ \bibinfo {pages} {2571} (\bibinfo {year}
  {1992})}\BibitemShut {NoStop}%
\end{thebibliography}%

\newpage

\begin{figure*}[htbp]
 \centering
 \includegraphics[width=14cm]{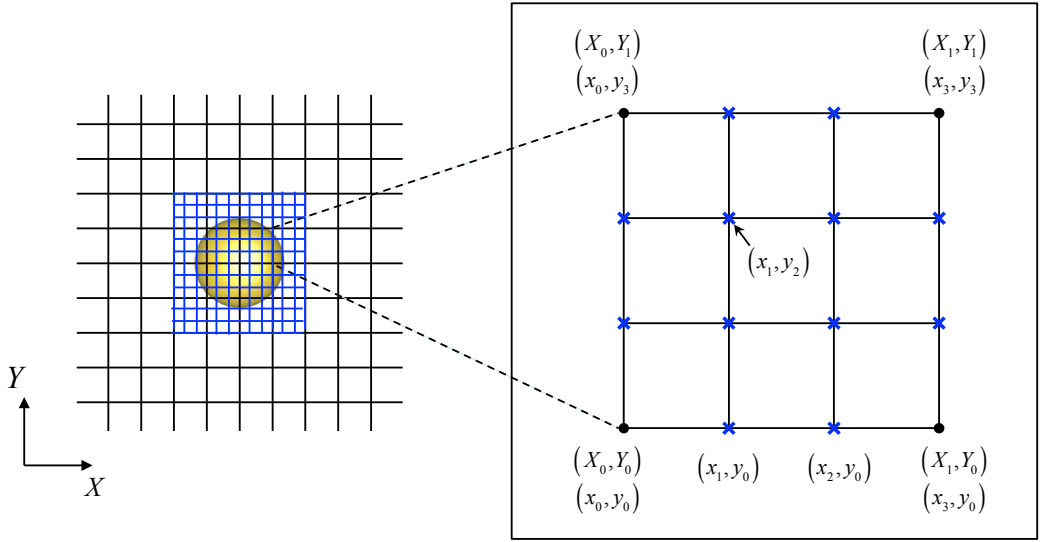}
 \caption{A two-dimensional schematic figure of the linear interpolation in case of $n$ = 3. The black dot (blue cross) marks in the right side figure represent the points on the 
coarse (interpolated) grid. The value of the Hartree potential on the interpolated grid $\left( {{x_1},{y_2}} \right)$ $\equiv$ (${s_x} = 1$, ${s_y} = 2$) is estimated by the four nearest points in two dimensional case as follows : ${V_\text{H}}\left( {{x_1},{y_2}} \right) = \frac{2}{3} \cdot \frac{1}{3}{V_\text{H}}\left( {{X_0},{Y_0}} \right) + \frac{1}{3} \cdot \frac{1}{3}{V_\text{H}}\left( {{X_1},{Y_0}} \right) + \frac{2}{3} \cdot \frac{2}{3}{V_\text{H}}\left( {{X_0},{Y_1}} \right) + \frac{1}{3} \cdot \frac{2}{3}{V_\text{H}}\left( {{X_1},{Y_1}} \right)$.
 }
 \label{Figure1}
\end{figure*}

\begin{figure*}[htbp]
 \centering
 \includegraphics[width=11cm]{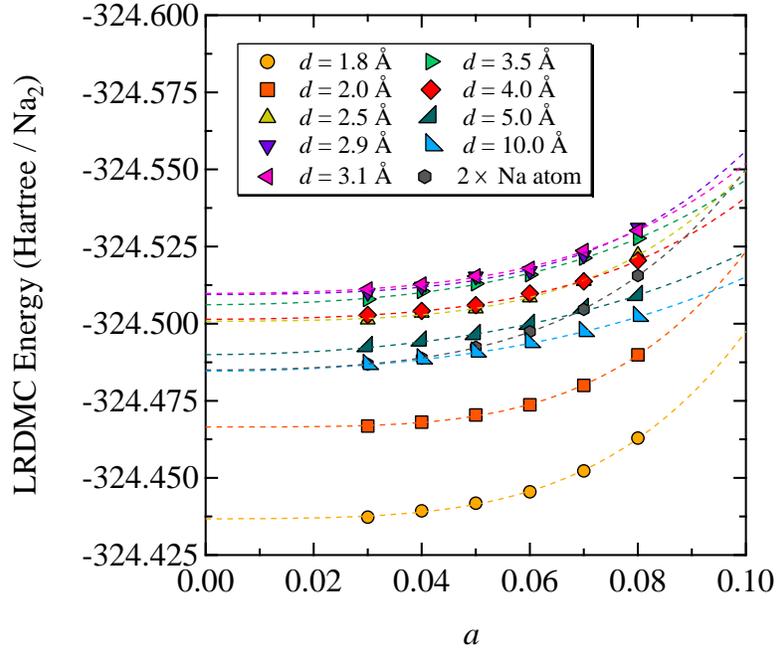}
 \caption{Total-energy vs $a$ in the LRDMC calculations of the sodium dimer. The error bars are within the markers.}
 \label{Figure2}
\end{figure*}

\begin{figure*}[htbp]
 \centering
 \includegraphics[width=12cm]{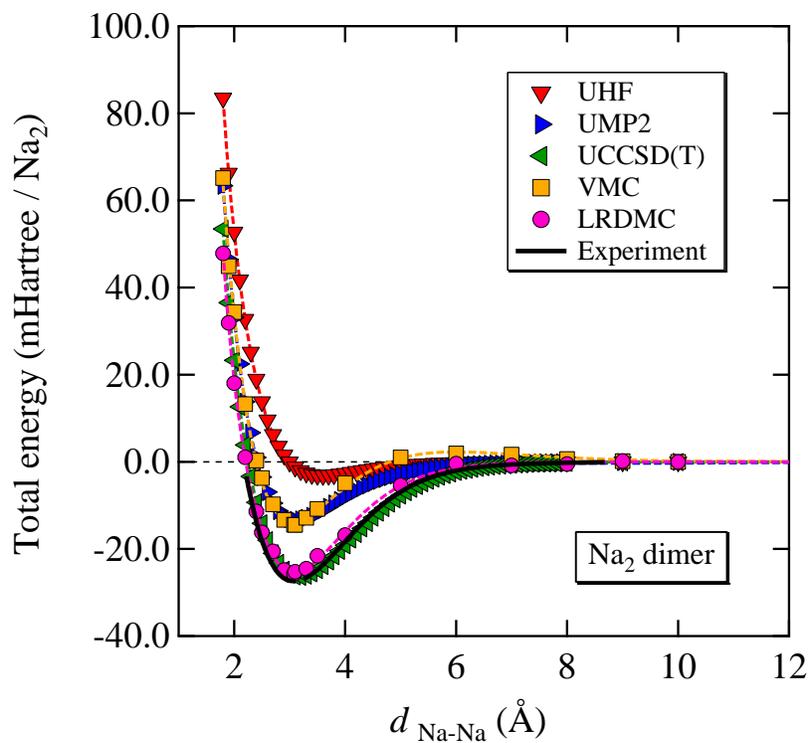}
 \caption{PESs of the sodium dimer. The broken lines show the obtained MS functions. The solid line shows the experimental values cited from Ref.{~\onlinecite{1983VER}}. The error bars are within the markers.
 }
 \label{Figure3}
\end{figure*}

\begin{figure*}[htbp]
 \centering
 \includegraphics[width=12cm]{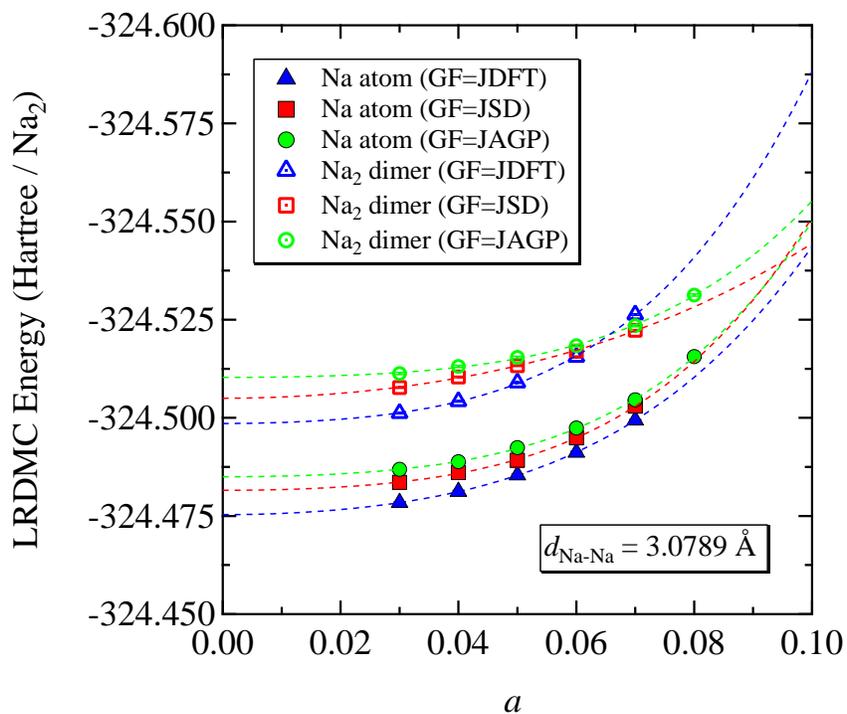}
 \caption{Total-energy vs $a$ in the LRDMC calculations of the sodium dimer at the experimental equilibrium distance ($d_\text{Na-Na}$ = 3.0789 \AA). The error bars are within the markers. GF denotes the guiding function.}
 \label{Figure4}
\end{figure*}

\begin{figure*}[htbp]
 \centering
 \includegraphics[width=17.5cm]{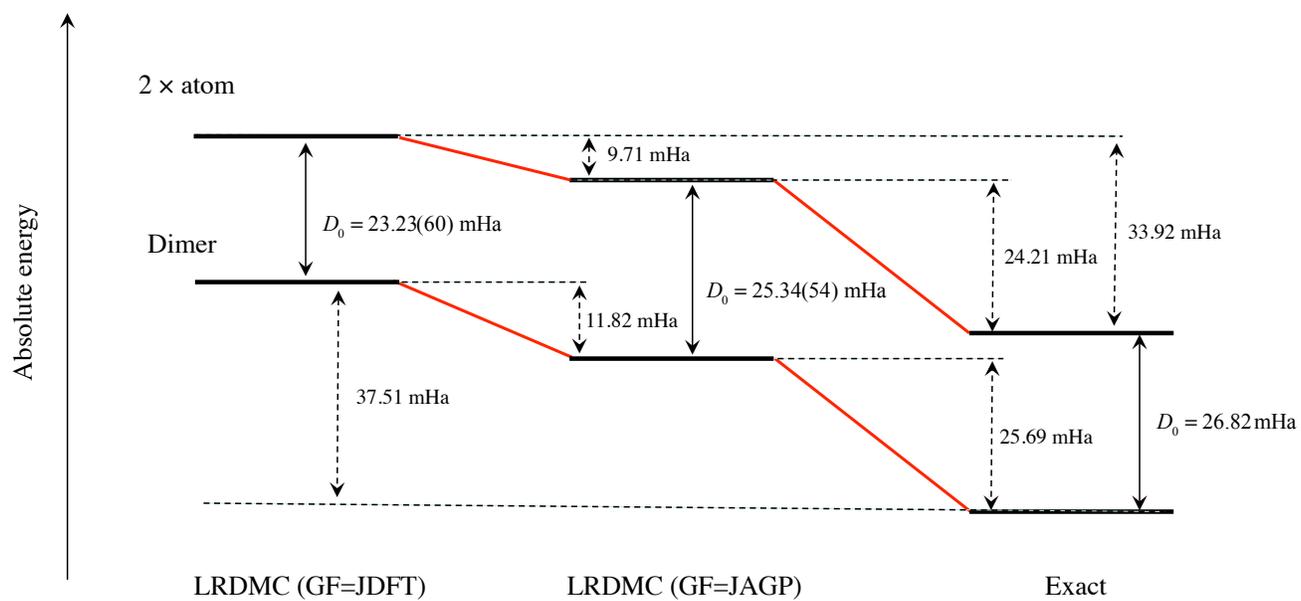}
 \caption{Diagram of the absolute energies of the sodium atom and dimer. The nodal surface errors and the binding energies are also shown. The exact energy of the sodium atom and the exact binding energy are cited from Ref.{~\onlinecite{1993CHA}} and {~\onlinecite{2013HUB}}, respectively. The exact energy of the sodium dimer was calculated these values.}
 \label{Figure5}
\end{figure*}


\newpage

\begin{center}
\begin{table*}[htbp]
\caption{\label{Table1} Ground state energies of the sodium atom obtained by HF, VMC, and LRDMC, and the estimated exact energy. GF denotes the guiding function.}
\vspace{3mm}
\begin{tabular}{c|ccc}
\Hline
Grid & Method & Energy (Ha) & Correlation (\%) \\ 
\Hline
\multirow{6}{*}{Reference}
& HF & -161.8589 {\footnote{Reference{~\onlinecite{1992SCH}}.}} & 0 \\
& VMC-JSD & -162.20717(33) {\footnote{Reference{~\onlinecite{2006BUE}}.}} & 88.0(1) \\
& VMC-JAGP & -162.1434(7) {\footnote{Reference{~\onlinecite{2003CAS}}.}} & 71.9(2) \\
& DMC (GF=JAGP) & -162.2370(1) {\footnote{Reference{~\onlinecite{2003CAS}}.}} & 95.6(3) \\
& DMC (GF=STO-HF) & -162.23966(22) {\footnote{Reference{~\onlinecite{2010NEM}}.}} & 96.2(2) \\
& Exact & -162.2546 {\footnote{Reference{~\onlinecite{1993CHA}}.}} & 100 \\ 
\hline
\multirow{3}{*}{}
& DFT-LDA & -161.4161 & - \\
& VMC-JDFT & -162.20442(21) & 87.32(5) \\
& VMC-JSD & -162.21423(16) & 89.80(4) \\
\multirow{1}{*}{Fine grid}
& VMC-JAGP & -162.22045(16) & 91.37(4) \\
\multirow{1}{*}{(0.02 Bohr)$^3$}
& LRDMC (GF=JDFT) & -162.23924(23) & 96.12(6) \\
\multirow{2}{*}{}
& LRDMC (GF=JSD) & -162.24017(57) & 96.35(14) \\
& LRDMC (GF=JAGP) & -162.24249(16) & 96.94(4) \\ 
\hline
\multirow{1}{*}{Fine grid}
& DFT-LDA & -165.5715 & - \\
\multirow{1}{*}{(0.05 Bohr)$^3$}
& VMC-JDFT & -154.554(11) & - \\
\hline
\multirow{1}{*}{Fine grid}
& DFT-LDA & Unstable & - \\
\multirow{1}{*}{(0.10 Bohr)$^3$}
& VMC-JDFT & Unstable & - \\
\hline
\multirow{3}{*}{}
& DFT-LDA & -162.9714 & - \\
& VMC-JDFT & -162.20151(21) & 86.58(5) \\
& VMC-JSD & -162.21474(17) & 89.93(4) \\
\multirow{1}{*}{Cubic interpolation}
& VMC-JAGP & -162.22079(16) & 91.46(4) \\
\multirow{1}{*}{(0.10 Bohr)$^3$ $+$ (0.01 Bohr)$^3$}
& LRDMC (GF=JDFT) & -162.23764(24) & 95.71(6) \\
\multirow{2}{*}{}
& LRDMC (GF=JSD) & -162.24078(20) & 96.51(5) \\
& LRDMC (GF=JAGP) & -162.24250(16) & 96.94(4) \\
\hline
\multirow{1}{*}{Cubic interpolation}
& DFT-LDA & -161.5461 & - \\
\multirow{1}{*}{(0.05 Bohr)$^3$ $+$ (0.01 Bohr)$^3$}
& VMC-JDFT & -162.20368(18) & 87.13(5) \\
\multirow{1}{*}{}
& LRDMC (GF=JDFT) & -162.23779(24) & 95.75(6) \\
\hline
\multirow{1}{*}{Cubic interpolation}
& DFT-LDA & -173.13140 & - \\
\multirow{1}{*}{(0.20 Bohr)$^3$ $+$ (0.01 Bohr)$^3$}
& VMC-JDFT & -162.18028(28) & 81.22(7) \\
\hline
\multirow{1}{*}{Linear interpolation}
& DFT-LDA & -174.67577 & - \\
\multirow{1}{*}{(0.20 Bohr)$^3$ $+$ (0.01 Bohr)$^3$}
& VMC-JDFT & -162.17665(29) & 80.30(7) \\
\Hline

\end{tabular}
\end{table*}
\end{center}

\begin{center}
\begin{table*}[htbp]
\caption{\label{Table2} Summary of the VMC-JAGP and LRDMC calculations of the sodium dimer.}
\vspace{3mm}
\begin{tabular}{c|c|c}
\Hline
$d_\text{Na-Na}$ (\AA) & $E_\text{VMC-AGP}$ (Ha/Na$_2$) & $E_\text{LRDMC}$ (Ha/Na$_2$) \\
\Hline
1.8 & -324.37708(34) & -324.43673(43) \\
1.9 & -324.39731(25) & -324.45269(45) \\
2.0 & -324.40784(33) & -324.46654(44) \\
2.2 & -324.42894(27) & -324.48354(46) \\
2.4 & -324.44191(22) & -324.49596(44) \\
2.5 & -324.44593(30) & -324.50076(44) \\
2.7 & -324.45197(25) & -324.50514(45) \\
2.9 & -324.45554(25) & -324.50944(44) \\
3.0789 & -324.45664(23) & -324.51033(44) \\
3.1 & -324.45668(21) & -324.50982(43) \\
3.3 & -324,45505(23) & -324.50911(44) \\
3.5 & -324.45303(29) & -324.50616(44) \\
5.0 & -324.44120(28) & -324.48999(43) \\
6.0 & -324.44028(28) & -324.48502(42) \\
7.0 & -324.44054(23) & -324.48538(42) \\
8.0 & -324.44156(20) & -324.48511(43) \\
9.0 & -324.44202(30) & -324.48443(42) \\
10.0 & -324.44219(30) & -324.48459(43) \\
2 $\times$ Na atom & -324.44158(32) & -324.48500(31) \\
\Hline
\end{tabular}
\end{table*}
\end{center}

\begin{center}
\begin{table*}[htbp]
\caption{\label{Table3} Summary of the obtained dissociation energies, equilibrium distances and harmonic vibrational frequencies in this work. Those obtained by previous ab initio calculations and the experimental values are also listed.}
\vspace{3mm}
\begin{tabular}{ccccccc}
\Hline
Method & $D_\text{e}$ (mHa) & $d_\text{eq}$ (\AA)& ${\omega _\text{e}}$ (cm$^{-1}$) \\
\Hline
UHF & 3.20 & 3.60 & 78.93 \\
UCCSD(T) & 26.49 & 3.179 & 154.79 \\
QCISD {\footnote{Reference{~\onlinecite{2007BEN}}. $D_\text{e}$ = 0.72193 (eV), $d_\text{eq}$ = 0.31813 (nm), and ${\omega _\text{e}}$ = 151.63 (cm$^{-1}$).}} & 26.53 & 3.181 & 151.63 \\
VMC & 13.58(10) & 3.050(6) & 155.5(1.7) \\
LRDMC & 25.28(43) & 3.083(11) & 163.4(3.4) \\
Full valence CI {\footnote{Reference{~\onlinecite{1993MAG}}. $D_\text{e}$ = 5892 (cm$^{-1}$), $d_\text{eq}$ = 5.83 (Bohr), and ${\omega _\text{e}}$ = 159.1 (cm$^{-1}$).}} & 26.85 & 3.09 & 159.1 \\
Experiment {\footnote{Reference{~\onlinecite{1995LIU}}. $D_\text{e}$ = 6022.6 (cm$^{-1}$), $d_\text{eq}$ = 5.82 (Bohr), and ${\omega _\text{e}}$ = 159.1 (cm$^{-1}$).}} & 27.44 & 3.08 & 159.1 \\
Experiment {\footnote{Reference{~\onlinecite{2013HUB}}. $D_\text{e}$ = 0.7298 (eV), $d_\text{eq}$ = 0.3079 (nm), and ${\omega _\text{e}}$ = 159.12 (cm$^{-1}$).}} & 26.82 & 3.079 & 159.12 \\
\Hline
\end{tabular}
\end{table*}
\end{center}






\begin{center}
\begin{table*}[htbp]
\caption{\label{Table4} Dissociation energies of the sodium dimer at the experimental equilibrium distance ($d_\text{Na-Na}$ = 3.0789 \AA). GF denotes the guiding function.}
\vspace{3mm}
\begin{tabular}{ccc}
\Hline
Method & GF & $E_\text{dimer-2atoms}$ (mHa) \\
\Hline
DMC & STO-HF & 23.87(57)\footnote{14.981$\pm$0.357 kcal/mol. See the supplementary material of ref.{~\onlinecite{2010NEM}}} \\
LRDMC & JDFT & 23.23(60) \\
LRDMC & JSD & 23.47(50) \\
LRDMC & JAGP & 25.34(54) \\
\Hline
\end{tabular}
\end{table*}
\end{center}
\begin{center}
\begin{table*}[htbp]
\caption{\label{Table5} The absolute energies of the sodium atom and dimer obtained by LRDMC (GF=JDFT, JAGP) and an experiment. The nodal surface errors (NS errors) of the absolute energy and the binding energies due to the error cancellations are also shown. The exact energy of the sodium atom and the exact binding energy are cited from Ref.{~\onlinecite{1993CHA}} and {~\onlinecite{2013HUB}}, respectively. The exact energy of the sodium dimer was calculated these values.}
\vspace{3mm}
\begin{tabular}{c|cc|cc|c}
\Hline
 & \multicolumn{2}{c|}{LRDMC (JDFT)} & \multicolumn{2}{c|}{LRDMC (JAGP)} & Experiment \\ 
\Hline
 & Energy (Ha) & NS error (mHa) & Energy (Ha) & NS error (mHa) & Exact energy (Ha) \\
\hline
2 atoms & -324.4753(48) & 33.92 & -324.4850(31) & 24.21 &-324.5092{\footnote{Reference{~\onlinecite{1993CHA}}.}} \\
Dimer & -324.4985(36) & 37.51 & -324.5103(44) & 25.69 & -324.5360	\\
\hline
$D_0$ (mHa) & \multicolumn{2}{c|}{23.23(60)} & \multicolumn{2}{c|}{25.34(54)} & 26.82{\footnote{Reference{~\onlinecite{2013HUB}}.}} \\
NS error (mHa) & \multicolumn{2}{c|}{3.59} & \multicolumn{2}{c|}{1.48} & - \\
\Hline
\end{tabular}
\end{table*}
\end{center}

\end{document}